\documentclass[prb,reprint,twocolumns,superscriptaddress]{revtex4-1}
\usepackage[utf8]{inputenc}
\usepackage[american,british]{babel}
\usepackage[T1]{fontenc}
\usepackage[pdftex]{graphicx}  
\usepackage{graphicx, xcolor}
\usepackage{dcolumn}
\usepackage{bm}
\usepackage{amsmath,amsthm,amssymb}
\usepackage{hyperref}
\usepackage{xcolor}
\hypersetup{colorlinks,bookmarksopen,bookmarksnumbered,
citecolor=magenta,
linkcolor=magenta,
pdfstartview=false,
urlcolor=magenta}
\usepackage{verbatim}
\usepackage{algorithm}
\usepackage{ulem}
\usepackage[noend]{algpseudocode}

\date{\today}

\begin{document}
\title{Measurement-induced criticality in (2+1)-d hybrid quantum circuits}

\author{Xhek Turkeshi}
\affiliation{The Abdus Salam International Centre for Theoretical Physics, strada Costiera 11, 34151 Trieste, Italy}
\affiliation{SISSA, via Bonomea 265, 34136 Trieste, Italy}
\affiliation{INFN, via Bonomea 265, 34136 Trieste, Italy}
\author{Rosario Fazio}
\affiliation{The Abdus Salam International Centre for Theoretical Physics, strada Costiera 11, 34151 Trieste, Italy}
\affiliation{Dipartimento di Fisica, Universit\`a di Napoli Federico II, Monte S. Angelo, I-80126 Napoli, Italy}
\thanks{On leave.}
\author{Marcello Dalmonte}
\affiliation{The Abdus Salam International Centre for Theoretical Physics, strada Costiera 11, 34151 Trieste, Italy}
\affiliation{SISSA, via Bonomea 265, 34136 Trieste, Italy}

\begin{abstract}
We investigate the dynamics of two-dimensional quantum spin systems under the combined effect of random unitary gates and local projective measurements. When considering steady states, a measurement-induced transition occurs between two distinct dynamical phases, one characterized by a volume-law scaling of entanglement entropy, the other by an area-law. 
Employing stabilizer states and Clifford random unitary gates, we numerically investigate square lattices of linear dimension up to $L=48$ for two distinct measurement protocols. For both protocols, we observe a transition point where the dominant contribution in the entanglement entropy displays multiplicative logarithmic violations to the area-law. We obtain estimates of the correlation length critical exponent at the percent level; these estimates suggest universal behavior, and are incompatible with the universality class of 3D percolation. \textbf{After the publication of the paper, an erratum has been pointed out. The latter is appended at the end of the paper. All the qualitative results remain the same}.
\end{abstract} 
\maketitle

\section{Introduction}
Entanglement plays a fundamental role in characterizing quantum many-body phenomena~\cite{Amico2007,Calabrese2009R,Eisert2010,Laflorencie2015}. A common setting where bipartite entanglement has attracted a great deal of attention is quantum quenches - i.e., the unitary time-evolution following a sudden change of the Hamiltonian parameters determining the system dynamics. Following a global quench starting from a generic low-entanglement (area-law) state, the von Neumann entropy of a given connected spatial partition grows linearly with time and relaxes to a value proportional to the partition volume (volume-law). 
Apart from remarkable exceptions, such as disorder-induced localized phases~\cite{Nandkishore2015,Abanin2019}, constrained quantum systems~\cite{Smith2017,Brenes2018,Surace2020,Sala2020,Russomanno2020} and long-range models~\cite{Schachenmayer2013,Buyskikh2016,Frerot2018,Liu2019,Lerose2020,Lerose2020B},  this trend is ubiquitous, as broadly documented by a wealth of theoretical studies~\cite{Calabrese2005,Calabrese2007,Rigol2008,Vengalattore2011,Kim2013,Mezei2017,Nahum2017,Bertini2019,Piroli2020,Bera2020,Chiara2006}. Thanks to conceptual and technological advances in cold atom and trapped ions systems, R\'enyi entanglement entropies of moderately large partitions are nowadays experimentally measurable~\cite{Daley2012,Islam2015,Kaufman2016,Elben2018,Brydges2019}.

Recently, a novel paradigm has been introduced in the study of entanglement dynamics, where unitary dynamics is interlayered with measurement operations~\cite{Cao2019,Li2019,Chan2019,Li2018,Skinner2019,Szyniszewski2019,Choi2019,Bao2019,Gullans2019,Gullans2019B,Tang2019,Jian2019,Zabalo2019,Zhang2020,Kuo2019,Gebhart2019,Snizhko2020,Goto2020,Rossini2020,Fan2020,Lavasani2020,Shtanko2020,Piqueres2020,Szyniszewski2020,Fuji2020,Chen2020,Ippoliti2020,Sang2020,Li2020,Lang2020}. 
This class of dynamics is an ideal testground to unveil the competition between local measurements and conventional Hamiltonian-type dynamics: for a low frequency of measurements, entanglement grows toward a volume-law, whilst a high rate of local measurement continuously collapses the state into a low-entanglement one.

The intermediate regime between area- and volume-law regimes has been extensively investigated in both random unitary circuits and Hamiltonian systems, with measurement paradigm varying from strong projective measurements, to weak continuous monitoring. Several studies have reported a second-order phase transition. Here, entanglement measures serve as order parameters, and universal behavior has been reported in the study of their finite-size scaling and their critical exponents. 
The specific case of hybrid random circuits (HRC) has been vastly investigated in (1+1)-d, where an underlying emergent conformal field theory (CFT) has been observed. 
The nature of this transition has been subject to debate. Motivated by numerical observations and analytical treatment, this critical point has been initially conjectured to lie in the same universality class of the 2D classical percolation theory transition~\cite{Li2019,Skinner2019}. However, more recent studies employing conformal field theory tools~\cite{Li2020} support that the transition in (1+1)-d hybrid circuits belongs to a different universality class than that of 2D percolation. 
Compared to the already rich (1+1)-d case, relatively little is instead known about their higher dimensional counterparts, where, even at equilibrium, entanglement properties are considerably different. For instance, the nature of a measurement-induced transition in two spatial dimensions could shed light on the relationship between HRC and percolation theories, and, potentially, give access to a new class of genuine out-of-equilibrium critical points.

In this work, we study the dynamics of (2+1)-d HRC. Using stabilizer states and Clifford unitary gates, we overcome known difficulties with more generic evolution protocol and reach extensive system sizes (square lattices of side up to $L=48$). 
We consider two measurement protocols, with rank-1 and rank-2 local projective measurements. In both cases, we find a volume-law phase at a slow rate of measurement, separated from an area-law phase at a high measurement rate via a measurement-induced transition (MIC). 
We perform a finite-size scaling (FSS) analysis to obtain accurate predictions of the correlation length critical exponent for the two cases. Our results suggest both critical points belong to the same universality class, which is distinct from that of percolation, similarly to the (1+1)-d case analyzed in Ref.~\onlinecite{Li2020}. 
This thesis is enriched by the presence of a violation of the area-law term in the entanglement entropy, that resembles those observed in Fermi liquids and $U(1)$ gauge fields coupled to fermionic matter~\cite{Wolf2006,Gioev2006,motrunich2008comparative,Swingle2010,Zhang2011,Swingle2016,Potter2014,Pouranvari2015}.

The rest of the paper is structured as follows. In Sec.~\ref{sec:model} we present the model and the observable under consideration, and we briefly discuss the tools implemented for numerical simulation. In Sec.~\ref{sec:crit} we present the numerical results for the entanglement entropy. Conclusions follows in Sec.~\ref{sec:conclusions}.

\begin{figure}[th]
\includegraphics[width=.9\columnwidth]{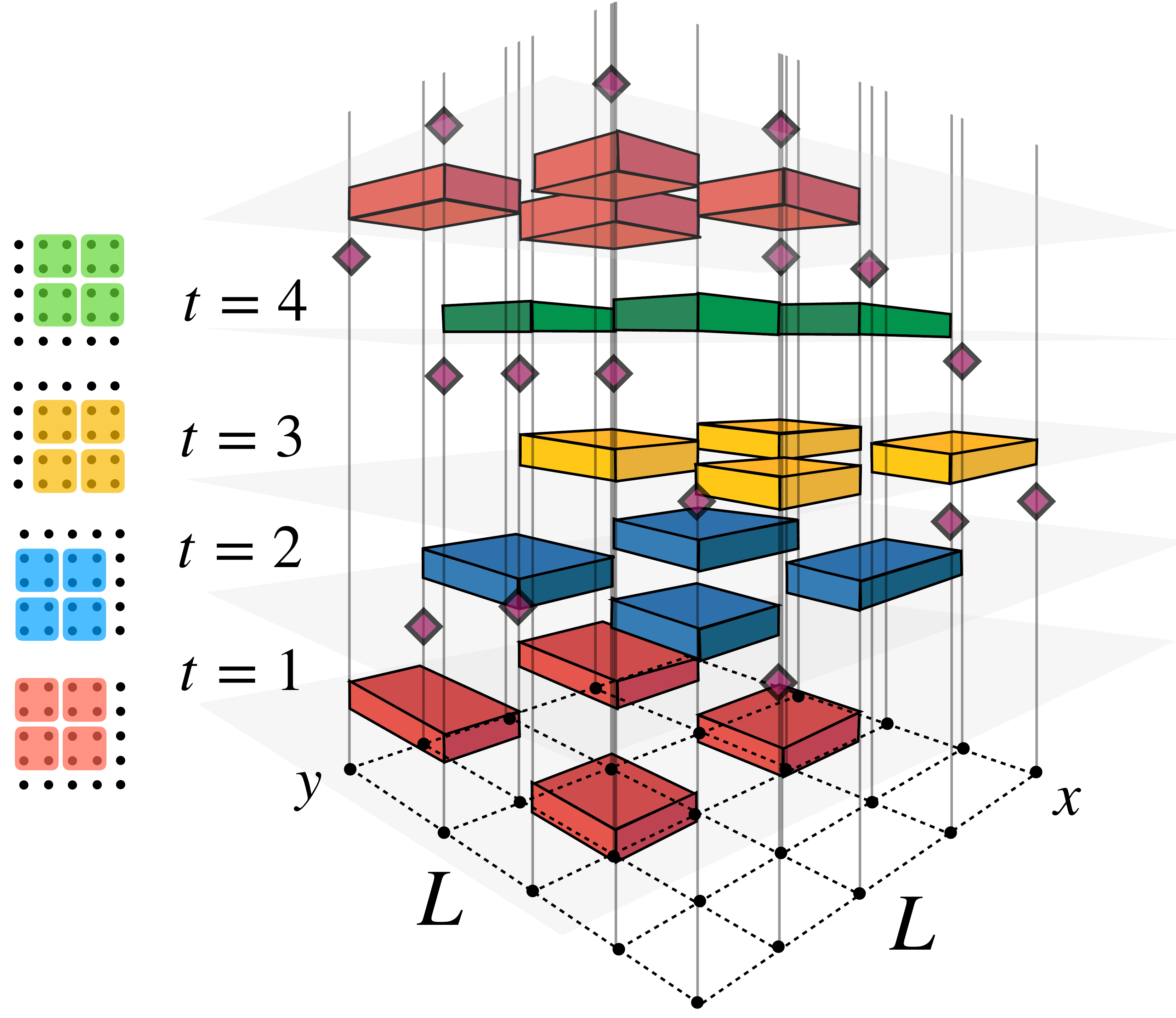}
	\caption{\label{fig:fig1v1} Graphical scheme of the system dynamics with measurement protocol in eq.~\eqref{eq:eq5.a.v1}. Spins are arranged on a square lattice ($x-y$ plane) of size $L\times L$. The time evolution is a stroboscopic sequence of random Clifford unitaries (colored rectangles) acting on plaquettes, and layers of random local projections (diamond symbols) acting on single spins. Different colors identify different unitary layers (see Eq.~\eqref{eq:eq1.v1}).  }
\end{figure}
\section{Model and observables}
\label{sec:model}

\subsection{System dynamics}

We consider a two-dimensional square lattice model of spin-1/2 qubits. The system is initialized in a low-entanglement state~\cite{foot1} and let evolve through a hybrid quantum circuit where unitary dynamics is alternated to layers of randomly picked local projective measurements (see cartoon in Fig.~\ref{fig:fig1v1}).

The unitary operations are given by random gates acting on four neighboring sites and structured in a brick-layer pattern. These gates have a periodic space-time pattern: depending on the value of the discrete time $t$, the operations are padded in the $x$ and $y$ directions. 
Given the elementary gate:
\begin{equation}
	\label{eq:eq1.v1}
	U(x,y,t) \equiv U_{(x,y),(x+1,y),(x,y+1),(x+1,y+1)}(t),
\end{equation}
each unitary layer is given by:
\begin{align}
	\label{eq:eq2.a.v1}
	U(t) &= \prod_{x=1}^{L_x/2}\prod_{y=1}^{L_y/2} U(2x-r_x(t),2y-r_y(t),t)\\
	r_x(t) &= \begin{cases}
		1,\quad \text{if } t\mod 4 = 1,2\\
		0,\quad \text{otherwise},
	\end{cases}\\
	r_y(t) &= \begin{cases}
		1,\quad \text{if } t\mod 4 = 0,1\\
		0,\quad \text{otherwise}.
	\end{cases}
\end{align}
The different shifts guarantees that the dynamics correlates all spins. The above operators act linearly on the state:
\begin{equation}
\label{eq:eq3v1}
	|\psi(t+1)\rangle= U(t) |\psi(t)\rangle,
\end{equation}
and generate entanglement throughout the system. 

\begin{figure}[th]
	\includegraphics[width=0.9\columnwidth]{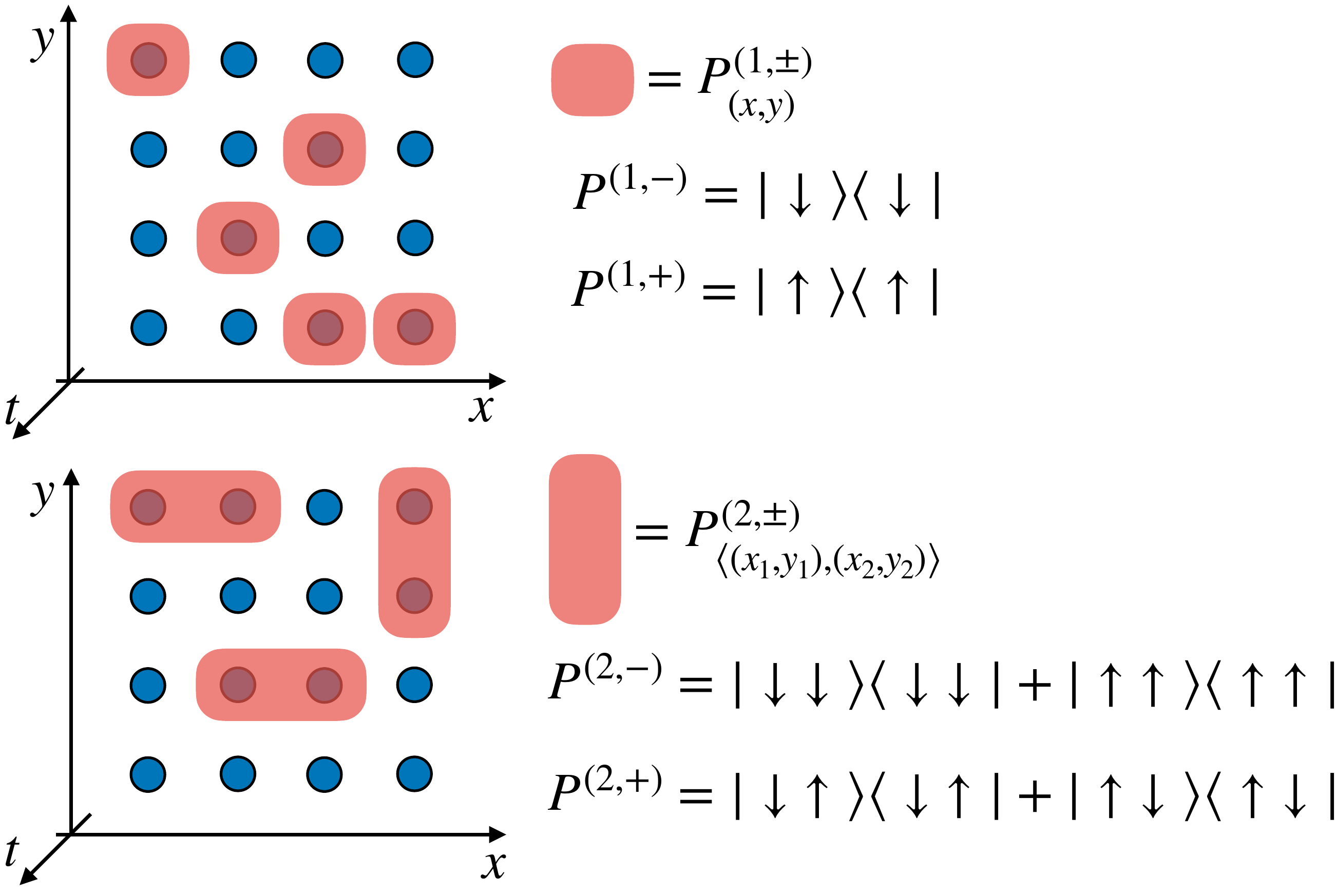}
	\caption{\label{fig:fig2v1} Local projection operator employed in the two dynamics. }
\end{figure}

Measurements are randomly picked with probability $p$ throughout the circuit. Given an evolution up to time $T$, for a square lattice of side $L$, the average number of measurements in the circuit is the fraction ${N_\textup{meas}=p L^2 T}$, where $0\le p\le 1$. 
These operations induce a non-linearity in the dynamics, as the wave function is renormalized after each collapse:
\begin{equation}
\label{eq:eq4v1}
	|\psi(t)\rangle \mapsto \frac{P^{\alpha}|\psi(t)\rangle}{||P^{\alpha}|\psi(t)\rangle||}.
\end{equation}
In the last equation, $\alpha$ is a label of the measurement type. In this paper we consider the following rank-1 and rank-2 projective measurement (see Fig.~\ref{fig:fig2v1}):
\begin{align}
\label{eq:eq5.a.v1}
	P^{(1)}_{(x,y)}& = \frac{1\pm \sigma^z_{(x,y)}}{2},\\
\label{eq:eq5.b.v1}
	P^{(2)}_{\langle (x_1,y_1),(x_2,y_2)\rangle} &= \frac{1\pm \sigma^z_{(x_1,y_1)}\sigma^z_{(x_2,y_2)}}{2}.
\end{align}
Here the single site measurement eq.~\eqref{eq:eq5.a.v1} acts on site ${(x,y)} $,  while eq.~\eqref{eq:eq5.b.v1} acts on neighboring sites ${(x_1,y_1)}$, ${(x_2,y_2)}$ and project the state onto a Bell pair.
Furthermore, the dynamics can be tailored conditionally on the measurement outcomes. In the present setting, we consider only unconditioned measurement layers, as the dynamic of entanglement for stabilizer states, is unaffected by the measurement outcomes.

For each circuit realization, we compute the entanglement entropy as a function of time. Given a bipartition of the system $A\cup B$, the entanglement entropy is defined as the Von Neumann entropy of the reduced density matrix $\rho_A(t) = \text{tr}_B |\psi(t)\rangle\langle \psi(t)|$:
\begin{equation}
\label{eq:eq6.v1}
	\mathbb{S}_A(\rho_A(t)) = -\text{tr}_A \rho_A(t) \log \rho_A(t).
\end{equation}
This quantity is an operational measure of entanglement, and in the present setting, serves as an order parameter characterizing distinct dynamical phases.
The latter are a consequence of the competing tendencies of the unitary evolution and the local projective measurements, whose balance is controlled by the rate $p$. 
A qualitative understanding of these dynamical phases is captured in the extreme limits~\cite{Li2018,Li2019,Skinner2019}. When ${p\sim 0}$, the evolution is largely unitary and the system is driven toward generic ("infinite temperature") wave-functions. For any given basis, the number of components of the Hilbert space required to sensibly capture this stationary regime scales exponentially with the subsystem volume, thus resulting in an extensive entanglement entropy $\mathbb{S}_A \propto \text{vol}(A)$.
On the other hand, when measurements are frequent ${p\lesssim 1}$, the projections impede information spreading beyond arbitrary distant regions of the system. In this regime, spins are correlated on a length-scale proportional to the domain of the projective measurement. As a consequence, the stationary wave-function is localized in a smaller subspace, resulting in an entanglement entropy scaling with the area of the subsystem boundary $\mathbb{S}_A\propto \text{area}(\partial A)$.
Finally, entanglement entropy is able to capture also a transition point between the aforementioned volume-law and area-law phases (see Sec.~\ref{sec:crit}). 

As randomness enters the model in both unitary and measurement layers, we are interested in the average values over many realizations for the hybrid circuits~\cite{foot2}:
\begin{equation}
\label{eq:eq7.v1}
	S_A(p,L)= \overline{\mathbb{S}_A(\rho_A(t))}=-\overline{\text{tr}_A \rho_A(t) \log \rho_A(t)}
\end{equation}
where we denote as $\overline{B}$ the average of a given quantity $B$ over the ensemble of realizations. We note that, in defining $S_A(p, L)$, the order of average is important, as the entanglement entropy is a non-linear functional of the density matrix. In fact, the average density matrix $\Phi = \overline{\rho_A(t)}$ always presents a volume-law compatible with that of thermal systems~\cite{Li2018}.

\subsection{Stabilizer states and Clifford unitary gates}
Truly generic random evolution would involve gates drawn with Haar measure from the full unitary group (usually denoted Haar gates). However, the exponential scaling of the Hilbert space hinders classical computations beyond a few decades of spins. Despite the remarkable results in obtained (1+1)-d numerical investigations, Haar gates are inadequate to tackle (2+1)-d dynamical problems. 
Thus, in order to achieve large numerical simulations and have a consistent scaling analysis, we restrict our attention to stabilizer states with unitary gates drawn from the Clifford group. 
The Clifford group is an approximation to the Haar gates, which fully encode statistical properties up to the second moment (2-unitary design)~\cite{Gottesman1998}. Remarkably, entanglement entropy within either Clifford or Haar circuits present similar features. Nonetheless, we stress the approximation breaks down when considering more complex objects. Important examples are out-of-time correlation (OTOC) functions, as they are one of the hallmarks of ergodicity in quantum systems. Simulations in (1+1)-d Haar circuits present evidence of exponential growth in time of the OTOC (signal of chaotic behavior), while analogous computations for Clifford circuits result in trivial time-scaling~\cite{Khemani2018,Nahum2018B,Pollmann2018}. 

In order to maintain the paper self-contained, we conclude this section with a technical summary on the stabilizer formalism and on the Clifford gates (we refer for a general review to Ref.~\onlinecite{Gottesman1998,Nielsen2012}). For readers already familiar with such formalism, the rest of the section is hopefully useful to clarify notations. After general considerations, we recall the Gottesman-Knill theorem~\cite{Gottesman1998,Aaronson2004} and the Hamma-Ionicioiu-Zanardi theorem~\cite{Hamma2004,Hamma2005}. The former explains how polynomial classical computation resources are needed to simulate the HRC of interest, while the latter gives an efficient way to compute entanglement for stabilizer states.

Stabilizer states are vectors of the Hilbert space satisfying the condition:
\begin{equation}
	\label{eq:eq8.v1}
	O_i |\psi\rangle = +1\cdot |\psi\rangle,
\end{equation}
for some set of operators $O_i$ (for spin-1/2 systems, we anticipate here these are Pauli strings that will be discussed below). This set, under matrix multiplication, forms a group ${G=\{O_i\}}$. In principle, the vectors satisfying eq.~\eqref{eq:eq8.v1} form a vector space associated to the group $G$. 
However, if the number of generators of the group is equal to the number of sites $N_s = L^2$, a unique state (up to normalization) is fixed by the knowledge of $G$ (see Ref.~\onlinecite{Nielsen2012}). 
Stabilizer formalism has been largely discussed in the context of quantum error correction (see Ref.~\onlinecite{Nielsen2012} and reference therein), and have recently appeared~\cite{Li2019,Li2018,Skinner2019,Choi2019,Bao2019,Gullans2019,Gullans2019B,Lavasani2020,Ippoliti2020,Sang2020,Li2020,Lang2020} in tailored non-unitary quantum dynamics as they can be efficiently simulated. 

The key result behind the simulations of stabilizer states under the action of the Clifford group is the Gottesman-Knill theorem, which explains: (i) how unitary evolution affects stabilizer states, (ii) how projective measurements change the state within the stabilizer formalism. Let us briefly sketch the ideas behind this result.
Under unitary evolution, eq.~\eqref{eq:eq8.v1} holds for the evolved stabilizer:
\begin{equation}
	\label{eq:eq9.v1}
	O_i(t) |\psi(t)\rangle = +1\cdot |\psi(t)\rangle,\quad O_i(t) = U(t) O_i U^\dagger(t).
\end{equation}
In general, $O_i(t)$ is a linear combinations over exponentially many Pauli strings.  However a major simplification occurs when the unitary $U$ is drawn from the Clifford group. The latter is defined as the set of  unitary operations that map a Pauli string into a \textit{single} Pauli string. 
Since the number of stabilizers does not grow under Clifford gates, the knowledge of the system only requires keeping track of the evolution of $N_s$ stabilizers at each time step.
However, a Pauli string is totally given by a binary vector of exponents and a phase:
\begin{align}
%	\label{eq:eq10.a.v1}
	O_i & = e^{i\phi}(\sigma_1^x)^{v^x_1}(\sigma_1^z)^{v^z_1}(\sigma_2^x)^{v^x_2}(\sigma_2^z)^{v^z_2}\dots (\sigma_{N_s}^x)^{v^x_{N_s}}(\sigma_{N_s}^z)^{v^z_{N_s}}\nonumber\\
	\label{eq:eq10.v1}
		& \equiv (v^x_1,v^z_1,	v^x_2,v^z_2,\dots,v^x_{N_s},v^z_{N_s}|\phi).
\end{align}
As a consequence, the state evolution under Clifford circuits is encoded by a $N_s\times (2 N_s + 1) $ matrix. We shall neglect the phase, as it does not contribute to entanglement. Thus our final state is encoded in a $N_s\times (2 N_s)$ matrix with binary entries.

Throughout this paper, we consider the Clifford group $\mathcal{C}_n$ acting on $n=4$ sites. For an efficient algorithm on how to implement uniform peaking over the Clifford group we refer to Ref.~\onlinecite{Koenig2014}.

\begin{figure}[th]
\centering
	\includegraphics[width=\columnwidth]{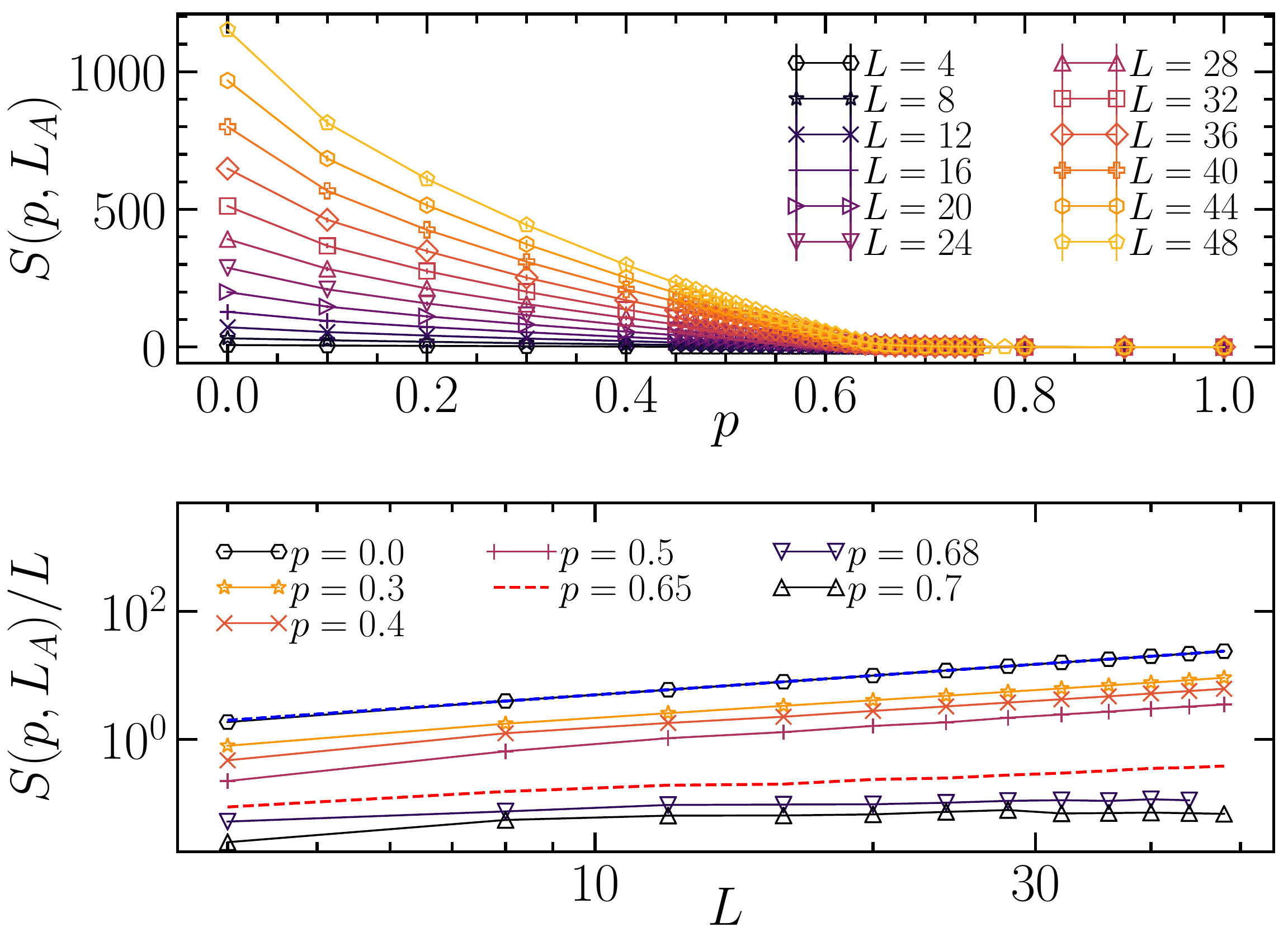}
	\caption{\label{fig:fig3v1} Entanglement for various linear system sizes $L$ and for various rates $p$, with $L_A = L/2$. We can see that the values $p< 0.65$ present a volume-law entanglement $S(p,L_A)\propto L_A^2$, while for values $p>0.65$ we have a quantum Zeno phase, with entanglement saturating to an area-law $S(p,L_A)\propto L$. The line $p_c= 0.65$ characterize the critical point, which exhibits a scaling $S(p_c,L_A)\propto L \log(L_A)$. The slope for a volume-law phase is provided in blue to guide the eye; this is quantitatively accurate for purely unitary dynamics. All error bars are smaller than the size of the symbols. }
\end{figure}

Projective measurement on Pauli string is less intuitive, but easy to implement. Let us consider a Pauli string $O_p$ we want to projectively measure on $|\psi\rangle$ a stabilizer state. Give its stabilizer group:
\begin{equation}
\label{eq:eq11.v1}
	G = \text{span}(O_1,O_2,\dots,O_k,O_{k+1},\dots,O_{N_s}),
\end{equation}
suppose that $[O_j,O_p]=0$ for $j\le k$ and $\{O_j,O_p\}=0$ for $j>k$ (either one holds for Pauli strings). The wave-function get mapped after measurement to:
\begin{equation}
	\label{eq:eq12.v1}
	|\psi\rangle_\pm \mapsto \frac{1\pm O_p}{2}|\psi\rangle.
\end{equation}
The Gottesman-Knill theorem states that the measured state expressed in term of the stabilizer group is given by:
\begin{align}
	G_\pm = \text{span}(&O_1,O_2,\dots,O_k,O_{k+1}\cdot O_{k+2},\dots\nonumber\\
	&O_{N_s-2}\cdot O_{N_s-1},O_{N_s-1}\cdot O_{N_s},\pm O_p).
\end{align}
If the outcome measure is of interest, for example in computing observable, the overall phase plays a relevant role as if affects expectation values. However, as already remarked, the phase is negligible for the entanglement computation and it is neglected in our computations.
In the case of study, $O_p = \sigma^z_i$, or $O_p=\sigma^z_i\sigma^z_j$. 

Lastly, entanglement entropy can be extracted directly for the binary matrix $N_s\times (2 N_s)$ encoding the state. Given a bipartition $A\cup B$ of dimension respectively $N_A$ and $N_B$, we extract the $N_s\times (2 N_A)$ matrix $G_A$ corresponding to the sites belonging in $A$. The Hamma-Ionicioiu-Zanardi theorem state the entanglement entropy for a stabilizer state is simply given by:
\begin{equation}
	\label{eq:eq13.v1}
	S_A(\rho_A) = \text{rank}(G_A) - N_A.
\end{equation}
As the rank is invariant under unitary operations, in (1+1)-d this gauge freedom has been used to fix a convenient ''standard'' form (clipped gauge). There, due to the simplicity of one-dimensional spin chains, a quasi-particle interpretation of the entanglement entropy has been given, as well as insights on the stabilizer length distribution~\cite{Li2019}.
We were not able to extend this picture to our (2+1)-d setting, thus we used SVD factorization to compute the rank in eq.~\eqref{eq:eq13.v1}.  
Let us conclude by remarking that all R\'enyi entropies for stabilizer states have the same values, implying a trivial spectrum of entanglement. This simply reflects the lack of complexity for higher-order cumulants for the Clifford circuits. 

\section{Entanglement dynamical phases and universal criticality}
\label{sec:crit}
We simulate the model in Sec.~\ref{sec:model} and compute the entanglement entropy averaged over $\mathcal{N}=10^4$ circuit realizations for each system size and each measurement rate considered. We consider periodic boundary conditions and consider bipartition for strips between $N_A = L\times L_A$ and $N_B=L\times (L-L_A)$. The latter choice allows us to isolate boundary contributions and neglect effects due to corners. We vary both $L$ and $L_A$ and store, after convergence is reached, the stationary value of the entanglement entropy $S(p,L_A)$. 

We have checked that, in the stationary regime, our results are independent of the initial state chosen (see Ref.~\onlinecite{Li2019} for similar results in (1+1)-d systems). 
Below we separately discuss the numerical results for rank-1 and rank-2 measurement considered (see Sec.~\ref{sec:model}). We find that for both protocols, a volume-law phase is separated by an area-law phase via a second-order phase transition (at a point $p_c$ which depends on the projector operator used). This point exhibits a universal behavior, in the sense that the computed correlation length critical exponents are compatible in the two cases within one error bar.

\begin{figure}[th]
\centering
	\includegraphics[width=\columnwidth]{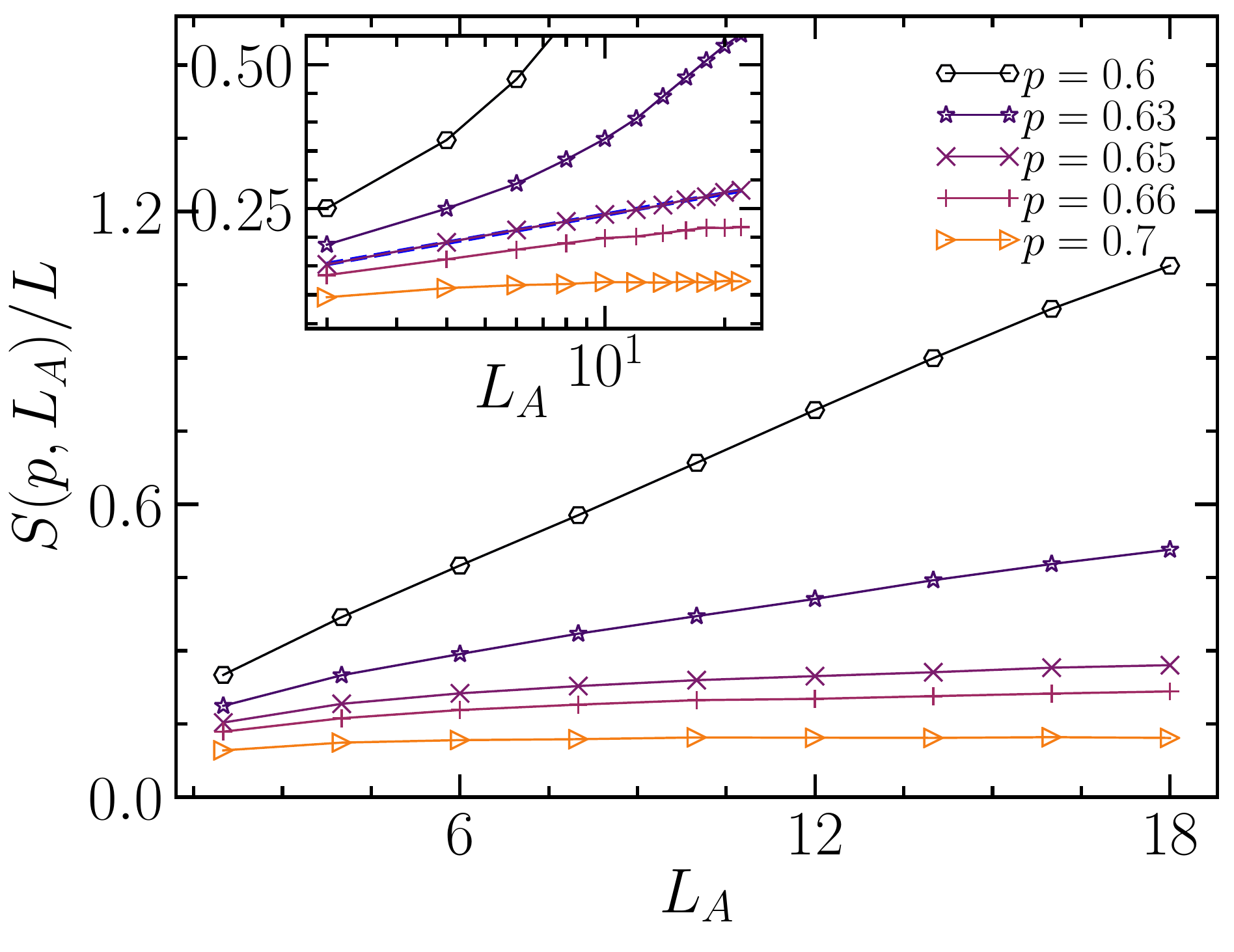}
	\caption{\label{fig:fig4v1} (Main) Stationary entanglement entropy near the critical point as a function of $L_A$. To spot the correct scaling, we divide the entanglement entropy by the uncut system size $S(p,L_A)/L$. Clearly one can distinguish three regimes: one super-logarithmic (volume-law), one logarithmic (critical point), and one sub-logarithmic (area-law). (Inset) Scaling of entanglement entropy close to the transition in logarithmic scale. To guide the eye, we plot the fitted $\log L_A$ for the ratio of interest. All error bars are smaller than the size of the symbols.}
\end{figure}

\subsection{Rank-1 measurements}
We consider the local projectors  $P_{\mathbf{i}}^{(1)}$ (cfr.~\eqref{eq:eq5.a.v1}). We simulate for various $p\in [0,1)$, expecting a volume-law average entanglement entropy for $p\simeq 0$ and an area-law for $p\lesssim 1$. The case $p=1$ is fine tuned, as the local projections applied to each sites project the state after each time step in a product state, hence not considered here.

In Fig.~\ref{fig:fig3v1} (top panel) we show the average entanglement entropy at half-system $S(p,L_A=L/2)$ for various values of system sizes $L$ and different measurement rates $p$. Here, errorbars are present, but are smaller than the size of the markers; thus they are not presented in the figures.
Since in two spatial dimensions the area-law is proportional to the entanglement cut length $L_A$, it is instructive to analyze $S(p,L_A)/L$, as this quantity saturates to a constant for an area-law phase and scale linearly with the system size in the volume-law phase.
In Fig.~\ref{fig:fig3v1} (bottom panel) we plot $S(p,L_A)/L$ for $L_A=L/2$. In the figure, it is possible to identify two distinct scaling regimes: For $p\lesssim 0.65$, the entropy increases linearly with the volume of the system. For $p\gtrsim 0.65$, after an initial growth for small sizes, the entropy saturates to a size-independent value. These results are expected from our previous discussion, except for the exact location of the critical point, which is extracted from a careful finite-size scaling analysis (presented below). 

Furthermore, in order to gain more information on the critical regime, we investigate the ratio $S(p,L_A)/L$ for a fixed system size, and varying the subsystem dimension. Specifically, we consider a lattice with side of length $L=48$, and consider a subsystem of dimension $N_A=L\times L_A$, with $L_A=4,8,\dots,L/2$. 
Also in this case, our simulations distinguish between the volume-law, area-law, and critical regimes, as presented in the Main panel of Fig.~\ref{fig:fig4v1}. To clearly characterize the critical line and its scaling, in the inset of Fig.~\ref{fig:fig4v1} we plot the results in a semi-logarithmic scale. Our data strongly support scaling at the transition of type $S(p_c,L_A)\propto L\ln L_A$. 

As remarked earlier, a correction of this kind has been, at present, observed only in Fermi liquid and in $U(1)$ gauge field coupled with matter. For Fermi liquids, the origin is hidden in the peculiar Fermi surface of the models~\cite{Gioev2006,Swingle2016,Potter2014}, and the entanglement scaling can be obtained analyzing the entanglement Hamiltonian. Similarly, the scaling of entanglement of critical spin liquids ($U(1)$ gauge fields coupled to fermions), has been observed numerically in Ref.~\onlinecite{Zhang2011}. The authors suggest this may be related to the fermionic matter of the theory, and their Fermi surface. 
In the present setting, the appearance of logarithmic corrections $\propto L\ln L_A$ is puzzling as: (i) the transition point is out-of-equilibrium, (ii) the system does not transparently have a Fermi surface. We postpone a discussion on the emergence of this scaling at the end of this section.

To access the correlation length critical exponent, we perform a finite-size scaling analysis (FSS) around the critical point. To compare with the literature in (1+1)-d HRC and the critical exponents of percolation theory, we use the scaling ansatz:
\begin{equation}
	\label{eq:eq14.v1}
	S(p,L_A)-S(p_c,L_A) = F((p-p_c) L_A^{1/\nu}).
\end{equation}
Specifically, given the scaling variables:
\begin{align}
    x_\textup{dat}(p_c,\nu)&=(p-p_c) L^{1/\nu},\\
    y_\textup{dat}(p_c)&=S(p,L)-S(p_c,L),
\end{align}
we implement polynomial fits for different degree polynomials and different subsets of system sizes. 
Given a fixed polynomial $P_m(x)$ of degree $m$ and given a subset of lengths $\{L_1,L_2,\dots,L_k\}$, the best fit is obtained minimizing the normalized least-square distance between the data and the polynomial computed on the scaling variable $x_\textup{dat}$:
\begin{equation}
    \label{eq:ffs}
    \varepsilon = \sqrt{\frac{\sum_{i} |y^2_\textup{dat}(i) - P^2_m(x_\textup{dat}(i))|}{\sum_{i} y^2_\textup{dat}(i)}}.
\end{equation}
\begin{figure}[th]
\centering
	\includegraphics[width=\columnwidth]{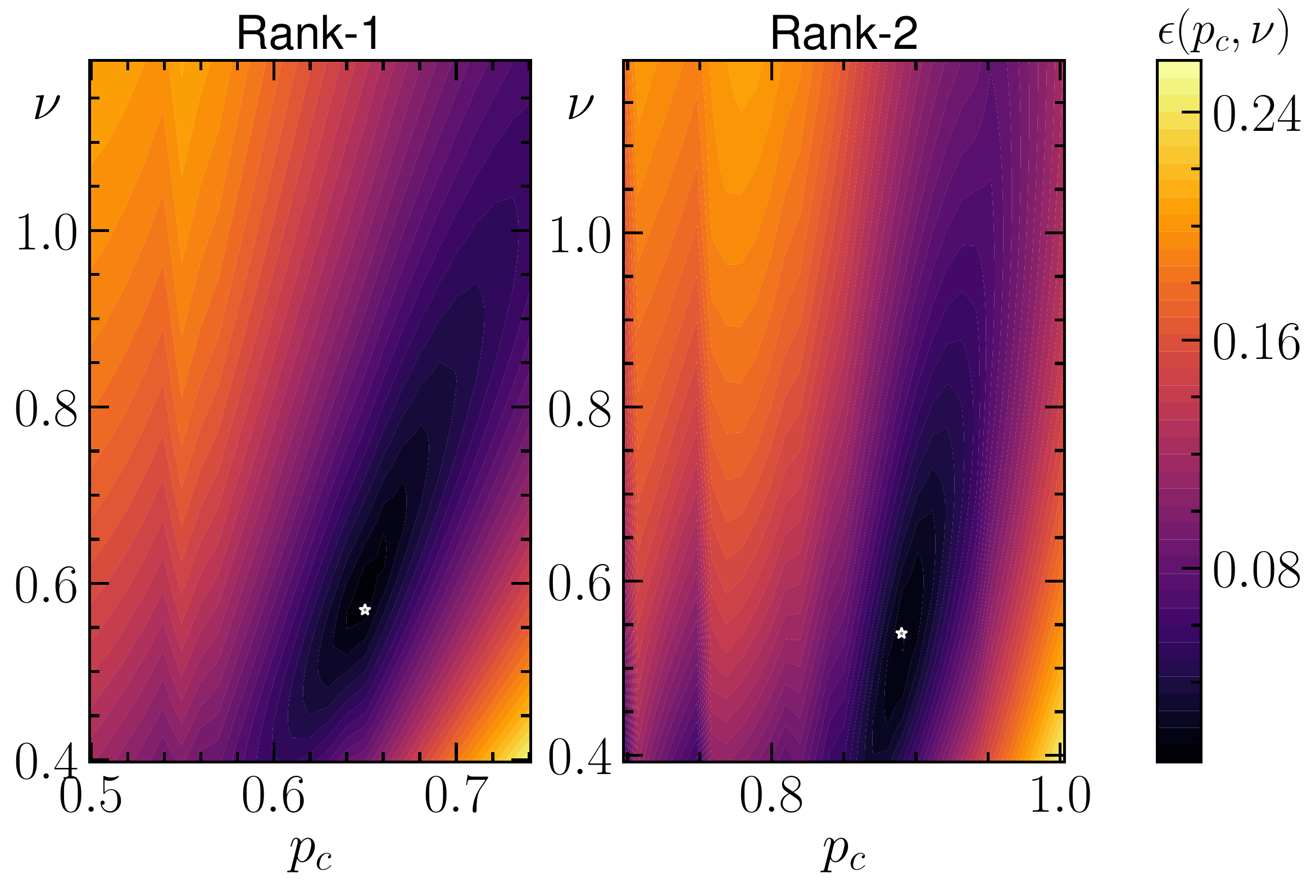}
	\caption{\label{fig:fssv2}Finite size analysis for the model of interest. The landscape of the residual is plotted for a suitable range of $\nu$ and $p_c$ considered in the FSS. The grey stars locate the optimal parameters. For the rank-1 HRC, this is at ${\nu=0.56}$, ${p_c=0.650}$, while for the rank-2 HRC it is at ${\nu=0.54}$, ${p_c=0.890}$. }
\end{figure}
Our final results are obtained averaging over  different values of the degree $m$ and different subsets of system sizes; similarly, the error is the propagated error. In Fig.~\ref{fig:fssv2} (left panel) we present the landscape of the residual for the optimal fit varying $\nu$ and $p_c$. 
The estimated critical parameters  ${\nu=0.56(1)}$ and ${p_c=0.650(5)}$, give us the data collapse in Fig.~\ref{fig:fig5v1}, presented in both linear and logarithmic scale.

\begin{figure}[th]
\centering
	\includegraphics[width=\columnwidth]{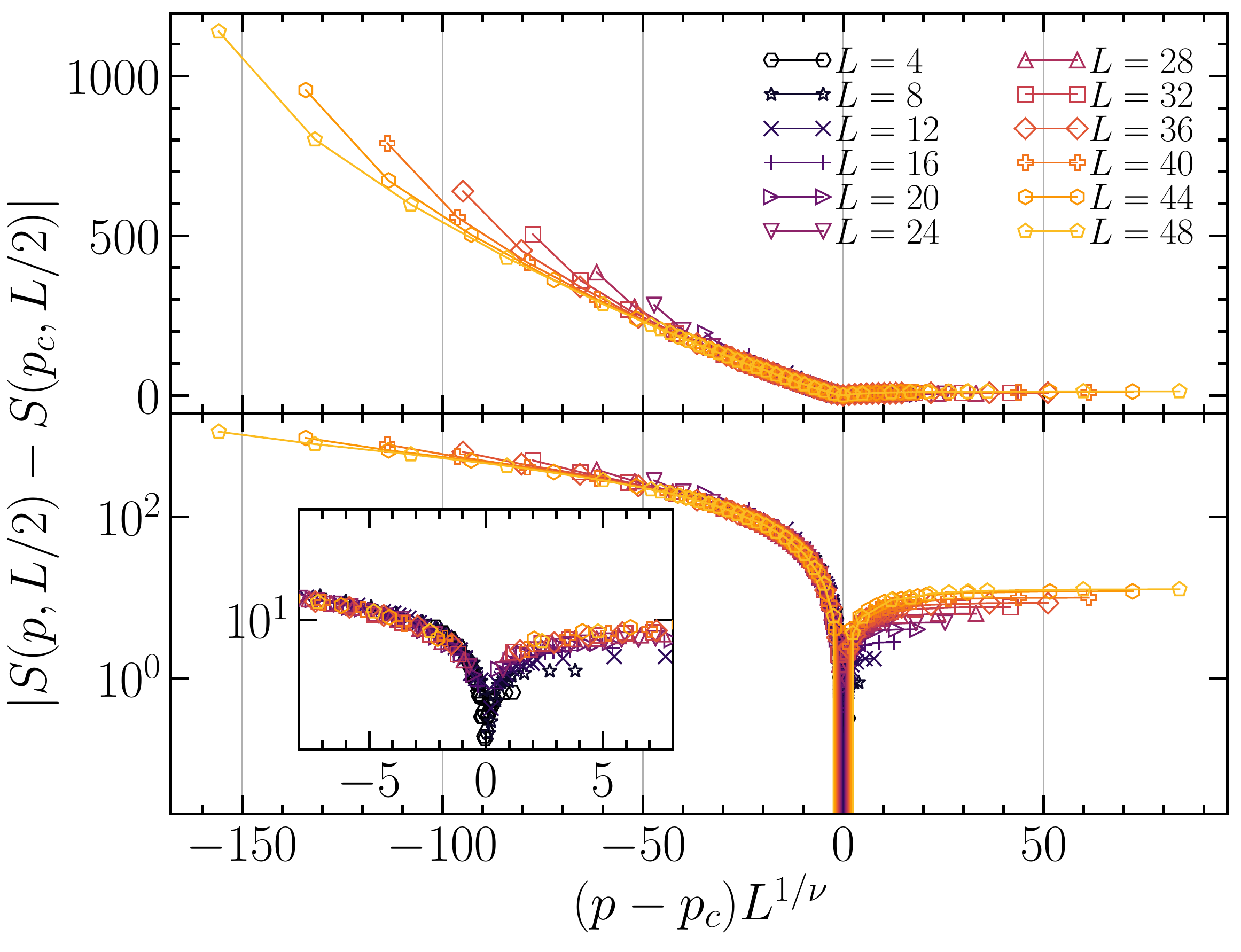}
	\caption{\label{fig:fig5v1} Data collapse for the hybrid circuit with rank-1 projective measurements. Here ${p_c = 0.650(5)}$ and ${\nu=0.56(1)}$. In the inset, we present a closer look on the critical point.}
\end{figure}

Let us conclude this subsection by comparing our results with the critical exponents of percolation theory. If a quantum-to-classical similarity has to hold in higher dimension (with respect to the results of (1+1)-d HRC), we should test our findings against 3D percolation on a cubic lattice. Here the correlation length critical exponent is $\nu^{3D}_\textup{perc}=0.877(1)$, more than $50\%$ different from our estimate. 
Consequently, this quantum critical point is sensibly different from the percolation one in the same dimension. 

\subsection{Rank-2 measurements}
\begin{figure}[th]
\centering
	\includegraphics[width=\columnwidth]{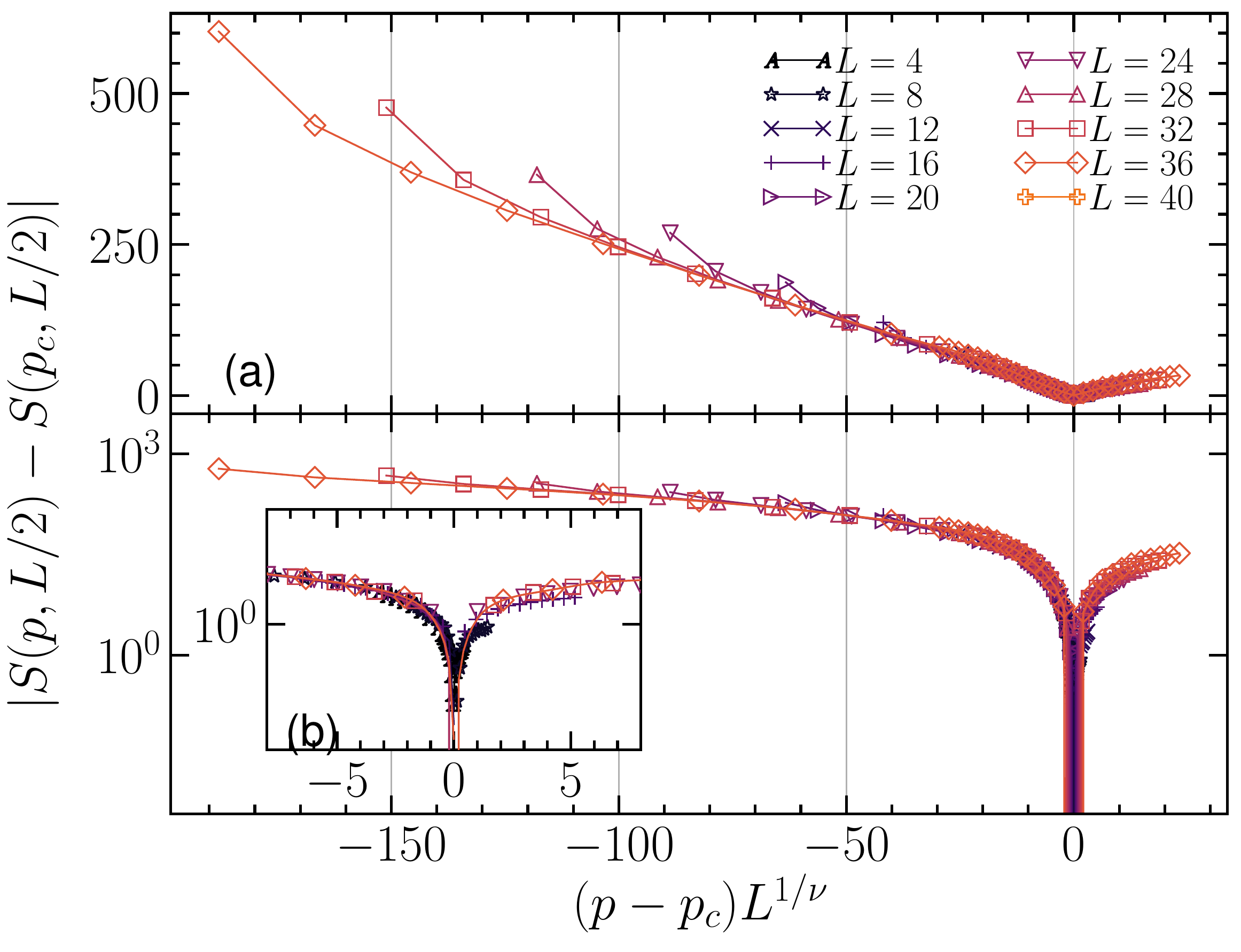}
	\caption{\label{fig:fig6v1} Data collapse of hybrid circuits with rank-2 projective measurements. Here $p_c = 0.890(3)$ and $\nu=0.54(1)$. The inset shows a closer look to the critical point. }
\end{figure}

These circuits have local projectors $P_{\mathbf{\langle i,\rangle j}}^{(2)}$ (cfr.~\eqref{eq:eq5.b.v1}). In this case, the measurement projects the neighboring qubits into a Bell pair. 
We perform a finite-size analysis analogous to the previous subsection. The landscape of the residual eq.~\label{eq:ffs} varying the parameters $p_c$ and $\nu$ is plotted in Fig.~\ref{fig:fssv2} (right panel). 
The critical point in this case is shifted to higher values (the estimated $p_c = 0.890(3)$). This is not a surprise, as the rank-2 projectors have less disentangling power than the correspondent rank-1 measurements, and the critical point is affected by changes in the microscopic physics of the system. Nevertheless, the universal information contained in the critical exponent is preserved, as we estimate $\nu = 0.54(1)$ (see Fig.~~\ref{fig:fig6v1} for the data collapse). 
This robust check confirms our previous analysis, in particular the distinction between this critical point and the percolation theory one. 

\subsection{Discussion and open questions}
The obtained numerical results leave us with open questions.  To which, if any, universality class do the measurement induced critical points belong? For $D>1$ hybrid quantum circuits, is there a classical effective model or any mean-field theory? Which is the origin of the area-law violation at the critical point? 
We conclude this section addressing these issues with speculative arguments based on our numerical observations.

Our estimate of the critical exponents (mutually compatible within $\%2$ error in both the considered models), suggest both critical points belong to the same universality class. However, we do not have enough information to fully characterize the nature of such a  universality class. 
A naive comparison with $D=3$ classical percolation theory rule out a quantum-to-classical analogy between entanglement and percolation in (2+1)-d circuits. In fact, our estimated critical exponent $\nu\simeq 0.55$ is incompatible from the $\nu_\textup{perc} \simeq 0.87$ of percolation theory in $3D$.

A key feature here is that the critical point exhibits an area-law violation. Such violation is common to gapless fermionic systems, such as free theories and Fermi liquids. It is however very unusual for spin systems: in these cases, such violations to area-law contributions are typically associated with the emergence of an underlying $U(1)$ gauge theory descriptions, with emergent fermionic excitations responsible for the logarithmic corrections~\cite{Grover2015}. We note that some classes of these gauge theories - directly connected to $\mathbb{C}P(N)$ models - have been reported to have critical exponents compatible with the one observed here~\cite{motrunich2008comparative}. This analogy in terms of entanglement scaling suggests that either the present critical regime has no-equilibrium analog, or that emergent fractionalization of quantum numbers might be taking place. The formulation of a rigorous statistical mechanics mapping as done in the (1+1)-d case, or the investigation of gauge-invariant quantum circuits may resolve this issue~\cite{Turkeshi2020}. 

From a complementary, microscopically oriented viewpoint, the logarithmic area-law violation we observe may be justified from the stabilizer size distribution. In one-dimensional systems, this quantity is defined in terms of the length of stabilizers, i.e. the distance between the edge Pauli matrices of a Pauli string. This has been related to entanglement entropy in Ref.~\onlinecite{Li2019,Chan2019}, where the authors deduce the following scaling in one spatial dimension:
\begin{equation}
\label{conj1}
S_A^{1+1d}(p,L)=\begin{cases}
\alpha(p)\log L + \beta(p)L& p<p_c,\\
\alpha(p)\log L& p=p_c, \\
\alpha(p)\log \xi & p>p_c.
\end{cases}
\end{equation}
Heuristically this argument extends to two spatial dimension, with the important caveat that here, for lattice models, a clear definition of stabilizer \textit{area} distribution is missing. (It is likely that corner effects may roughen a proper scaling limit). 
Nonetheless, from our numerical data we conjecture this is the case, and correspondingly the entanglement entropy behaves as:
\begin{equation}
\label{eq:conj2}
S_A^{2+1d}(p,L)=\begin{cases}
\tilde\alpha(p)L\log L + \tilde\beta(p)L^2& p<p_c,\\
\tilde\alpha(p)L\log L& p=p_c, \\
\tilde\alpha(p)L\log \xi & p>p_c,
\end{cases}
\end{equation}
with $\tilde\alpha$ and $\tilde\beta$ system-size independent.
We leave for future work elaborating a proper definition of the stabilizer area and its implication on the hybrid quantum circuits dynamics. 

\section{Conclusions and outlook}
\label{sec:conclusions}

We investigated the measurement-induced criticality in two-dimensional hybrid quantum circuits generated by Clifford random unitary gates. Our findings reveal that the entanglement transition separating area and volume-law phases present universal features: those are signaled by the correlation length critical exponent being insensitive to the choice of the measurement, and by the same functional form of the entanglement entropy at criticality, showing logarithmic violations of the area-law. This universality class is distinct from that of 3D percolation theory.

Concerning the nature of the critical point, at equilibrium, the entanglement scaling we report has been previously observed only in systems with fermionic excitations, such as Fermi liquids and $U(1)$ gauge theories coupled to fermionic matter. For future works, it may be interesting to characterize such entanglement transitions by both studying the interplay of unitary dynamics and measurements directly in gauge-invariant circuits, and elaborating generalizations of the stabilizer length distribution. 

From the computational side, additional insights may be gathered via the computation of other observables, such as the scaling of corner contributions in the entanglement entropy, and equal-time correlation functions. Regarding the latter, the challenge is to implement these quantities within the stabilizer formalism, thus preserving the technical advantage over full (Haar) quantum dynamics. Moreover, it would be interesting to seek the upper critical dimension of the system of interest, and in particular, if a "mean-field" regime can be captured by a classical statistical mechanics model. In fact, our estimate of the critical exponent $\nu\simeq 0.55$ is close to the mean-field limit $\nu_\textup{MFT}=1/2$ of statistical field theory. It is possible that already HRC in 3+1d saturate this limit, a fact that might be detectable already at modest system sizes due to its mean-field origin.

\begin{acknowledgments}
X.T. acknowledges useful discussions with S. Pappalardi. We thank G. Pagano, A. Russomanno, and S. Sharma for discussions.
X.T. and M.D. are partly supported by the ERC under the grant No 758329 (AGEnTh). M. D. is partly supported by the European Union's Horizon 2020 research and innovation program under grant agreement No 817482, and by the Italian Ministry of Education under the FARE programme. R. F. acknowledges a Google Quantum Research Award.
\end{acknowledgments}

\phantomsection
%%%%%%%%%%%%%%%%%%%%%%%%%%%%%%%%%%%%%%%%%%%%%%%%%%%%%%%%
%%%%%%%%%%%%%%%%%%%%%%%%%%%%%%%%%%%%%%%%%%%%%%%%%%%%%%%%
%%%%%%%%%%%%%%%%%%%%%%%%%%%%%%%%%%%%%%%%%%%%%%%%%%%%%%%%
%%%%%%%%%%%%%%%%%%%%%%%%%%%%%%%%%%%%%%%%%%%%%%%%%%%%%%%%

\clearpage
\onecolumngrid
\begin{center}
  \textbf{\Large Erratum: Measurement-induced criticality in (2+1)-dimensional hybrid quantum circuits}\\[.2cm]
\end{center}

\twocolumngrid

In our manuscript, we incorrectly reported that singular-value-decomposition (SVD) directly evaluates the entanglement entropy of a stabilizer state generated by the dynamics we are interested in. In fact, SVD only provides a rigorous upper bound to the entropy~\footnote{In 1D, results obtained via analyzing such upper bounds up in chains with up to $L=128$ spins return critical properties that are the same (within error bars) of those obtained via Gaussian elimination.}. We are grateful to Y. Li and M. P. A.  Fisher for correspondence that elucidated this aspect. 

Specifically, in the published version of the paper we compute entanglement of stabilizers state by means of the Hamma-Ionicioiu-Zanardi theorem. This requires the computation of the matrix rank. SVD computes the value of the rank in the field of real numbers $\mathbb{R}$, while, due to the algebraic structure of the stabilizer group, we should have computed the rank for the field $\mathbb{F}_2$.

We observe that any matrix $A$ with binary elements $a_{ij}=0,1$ satisfy:
\begin{equation}
\label{eq:1}
	\text{rank}_{\mathbb{F}_2} A \le \text{rank}_{\mathbb{Q}} A=  \text{rank}_{\mathbb{R}} A.
\end{equation}
In the previous equation, $\mathbb{F}_2$ is the finite field over $\mathbb{Z}_2$, while $\mathbb{Q}$ is the field of rational numbers.

Thus, in order to obtain the correct results for the models described in the published version of the paper, we need an exact computation of the rank over $\mathbb{F}_2$. This computation is done by Gaussian elimination, which works for any field $\mathbb{K}$. The algorithm scales as $\mathcal{O}(N_s^3)$ with $N_s\propto L^2$ the number of spins.

We have repeated all simulations and extracted the entanglement entropy using Gaussian elimination. All the corresponding results are presented below. The three main findings of our work \textit{are confirmed}, namely: 
\begin{enumerate}
	\item There is a universal critical behavior for 2+1D hybrid quantum circuits;
	\item The critical point shows a multiplicative logarithmic correction to the area law entanglement entropy;
	\item The universality class is different from that of 3D percolation.
\end{enumerate}

The critical values of $p_c$ and $\nu$ were instead not correct for both rank-1 and rank-2 measurement schemes. The correct values are $p_c=0.54(1), \nu = 0.67(1)$ and $p_c=0.84(1), \nu = 0.68(1)$ for rank-1 and rank-2 measurement schemes, respectively. Thus, compared to the previous results, the critical exponent $\nu$ for both model is 20\% different from the values reported previously, while the critical points $p_c$ are offset of 20\% for rank-1 measurements and of 6\% for rank-2 measurements.

\begin{figure}
	\includegraphics[width=\columnwidth]{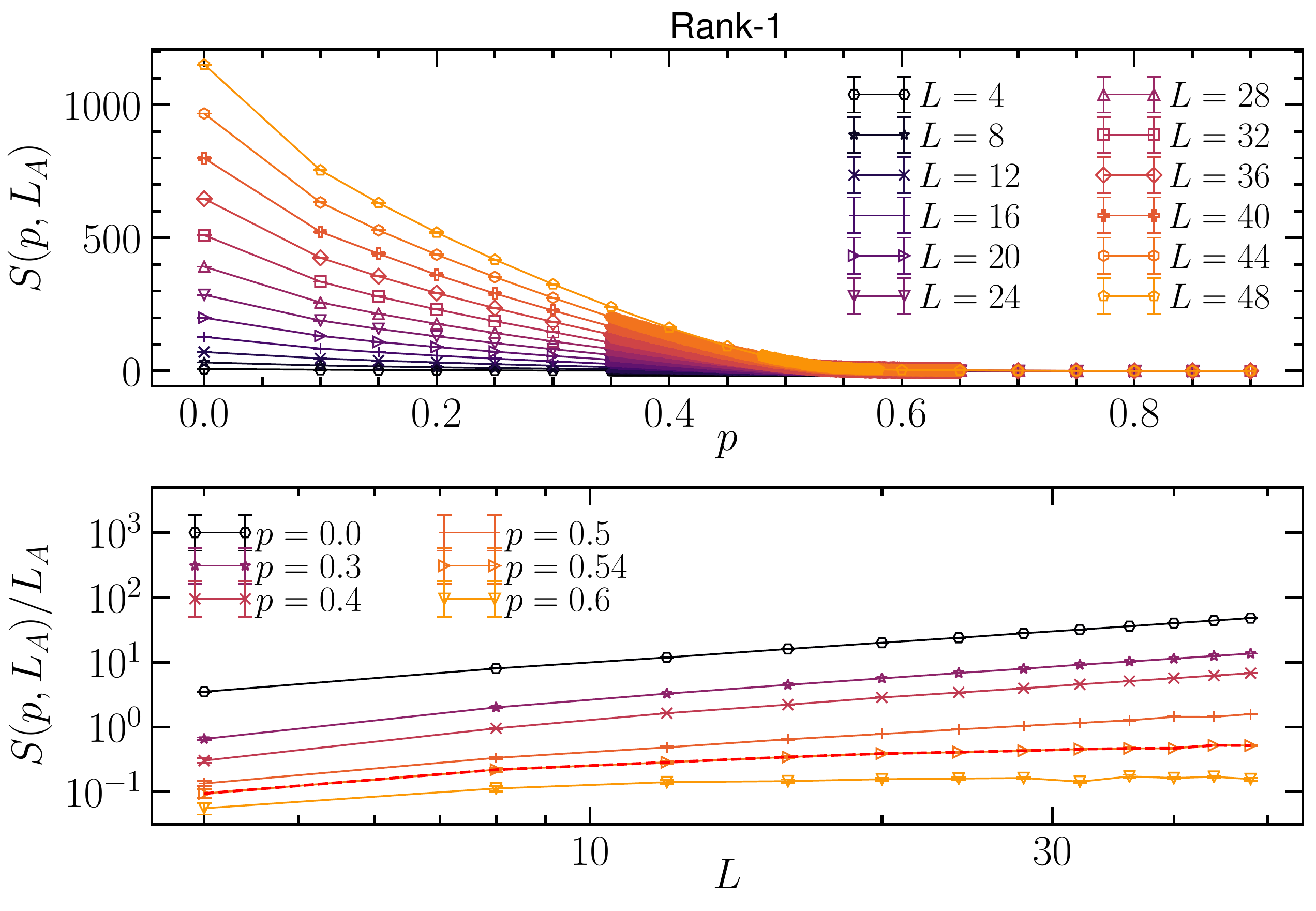}
	\includegraphics[width=\columnwidth]{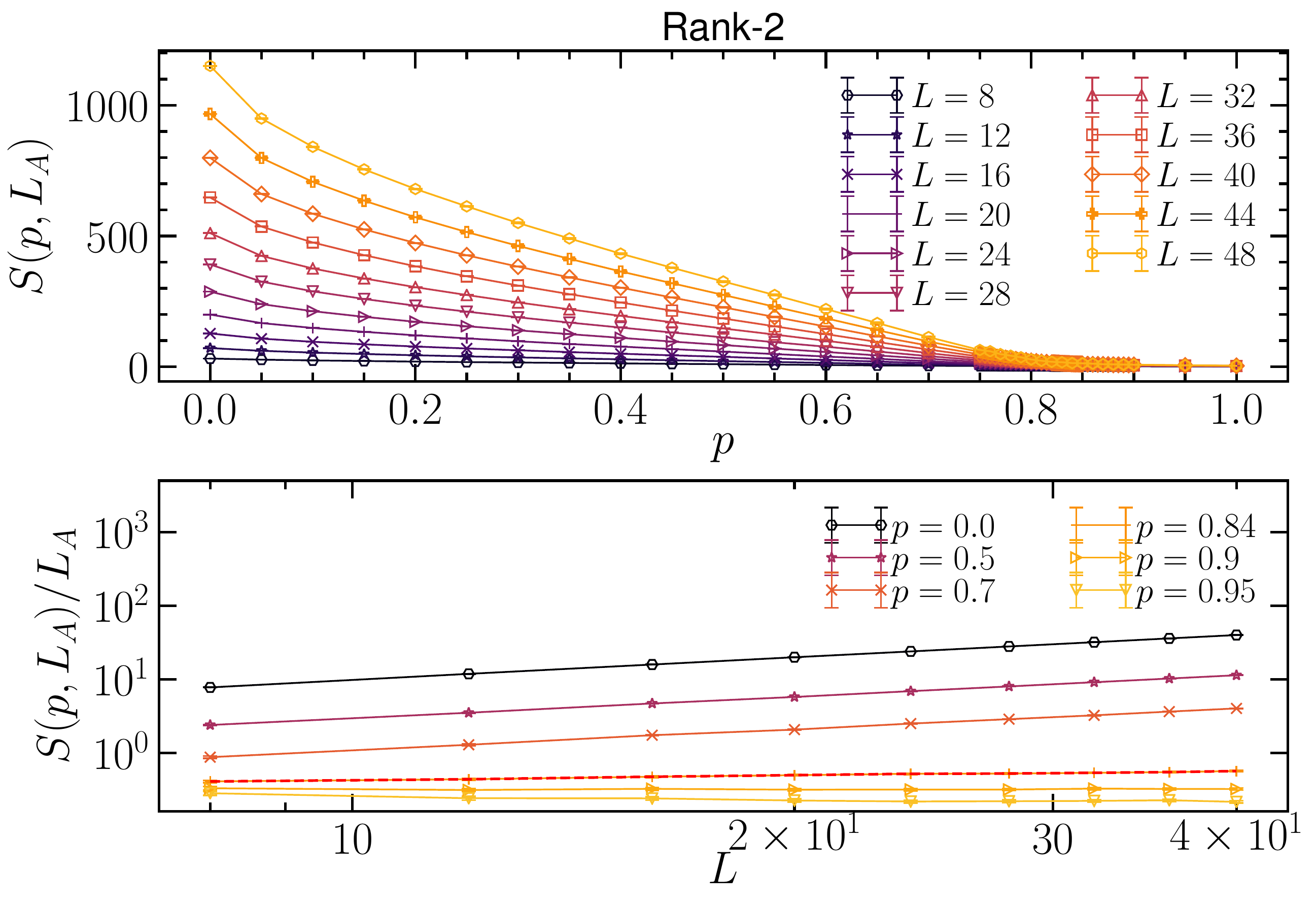}
	\caption{\label{fig1} Bare data for the hybrid random circuits entanglement entropy computed through gaussian elimination. The two upper panels pertain the rank-1 measurements, while the lower ones the rank-2 measurements. The red line characterize the critical line in both the models.  }
\end{figure}

\begin{figure}[ht!]
	\includegraphics[width=1.\columnwidth]{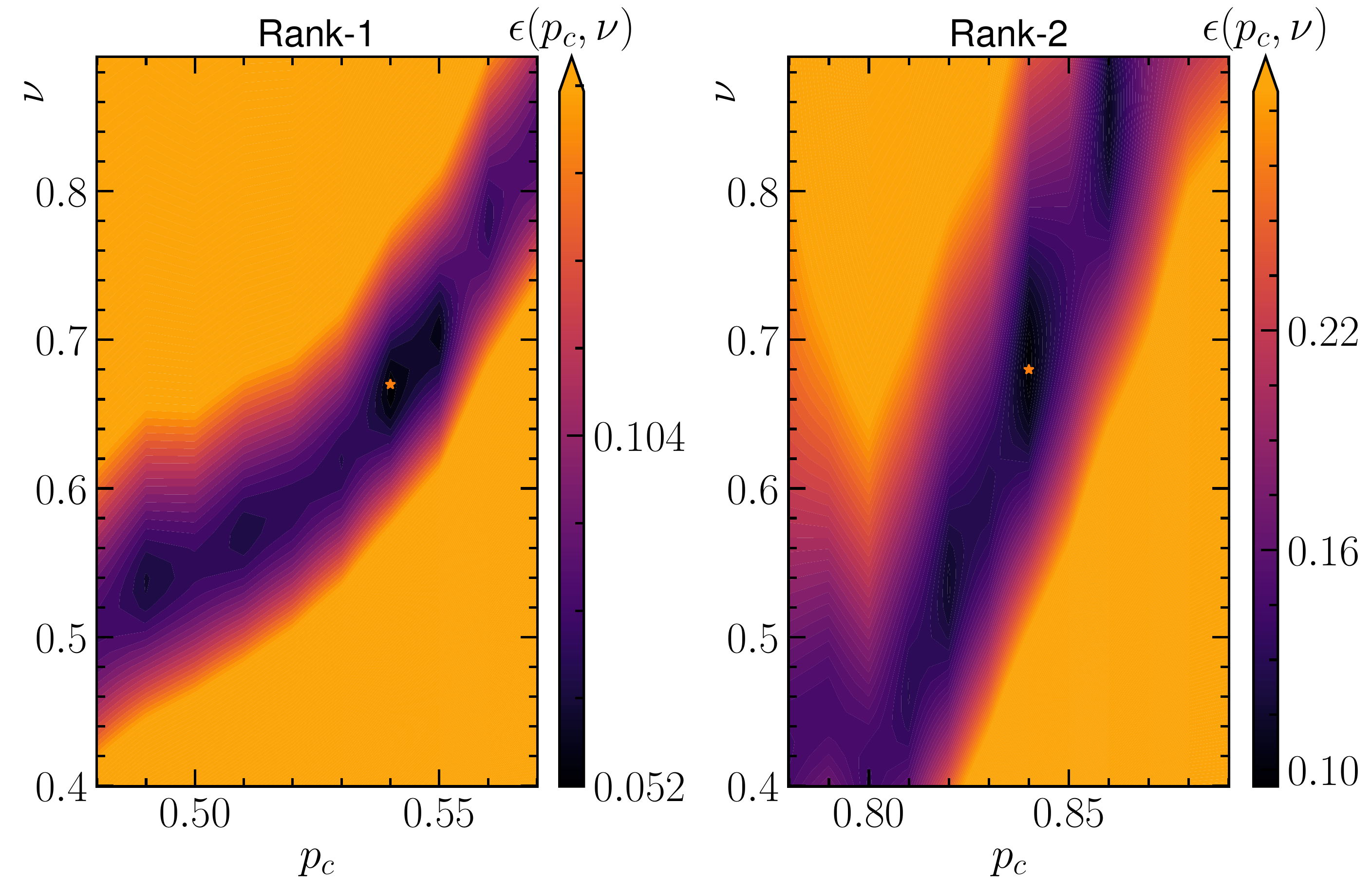}
	\caption{\label{fig2} Finite size scaling results for both the model. The procedure is the one presented in the published version of the paper. The updated results show a different estimate for the both the considered model with respect to the published results. In particular, the rank-1 model (left panel) gives $p_c=0.54(1)$, $\nu=0.67(1)$. The rank-2 model (right panel) gives $p_c=0.84(1)$, $\nu=0.68(1)$. For convenience, the respective critical values are pointed through a star in the figures.}
\end{figure}

The compatibility of the critical exponent for both the considered hybrid random circuit models within 2\% of error suggests universality holds for the corrected data. This exponent is $>30\%$ different from that of 3D percolation ($\nu_\textup{perc}^{3D}=0.87(1)$), suggesting the 2+1D HRC belong to a different universality class.

We conclude by pointing out that, despite the SVD gives an upper bound of our data, it present hint of universality (see published version of the paper). 
It is possible that the SVD is capturing different features of the model of interest, or some specific limit of it. We leave further investigation on its relationship to Clifford hybrid random circuits for future work.

\twocolumngrid
\begin{figure}[h!]
	\includegraphics[width=0.95\columnwidth]{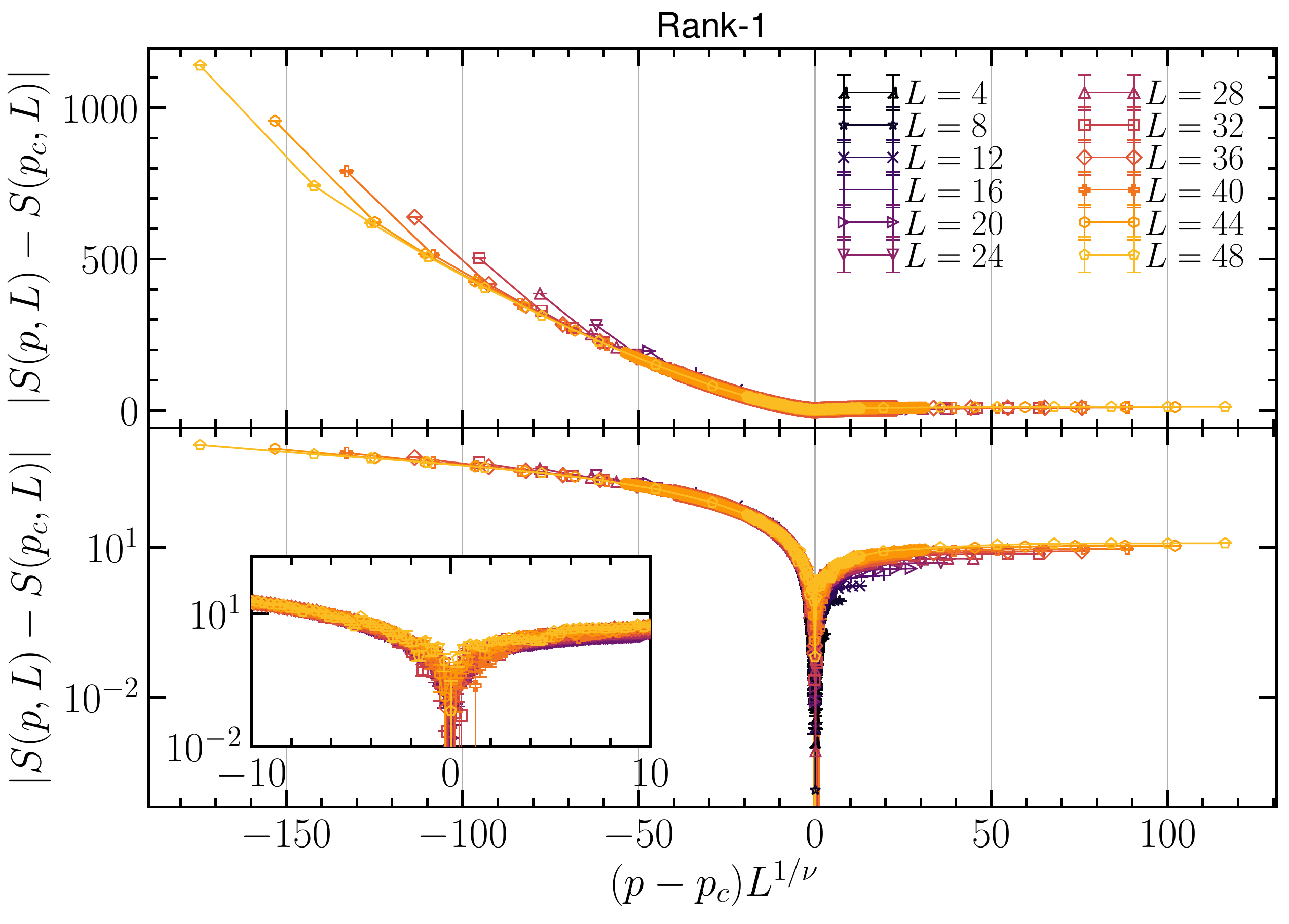}
	\includegraphics[width=0.95\columnwidth]{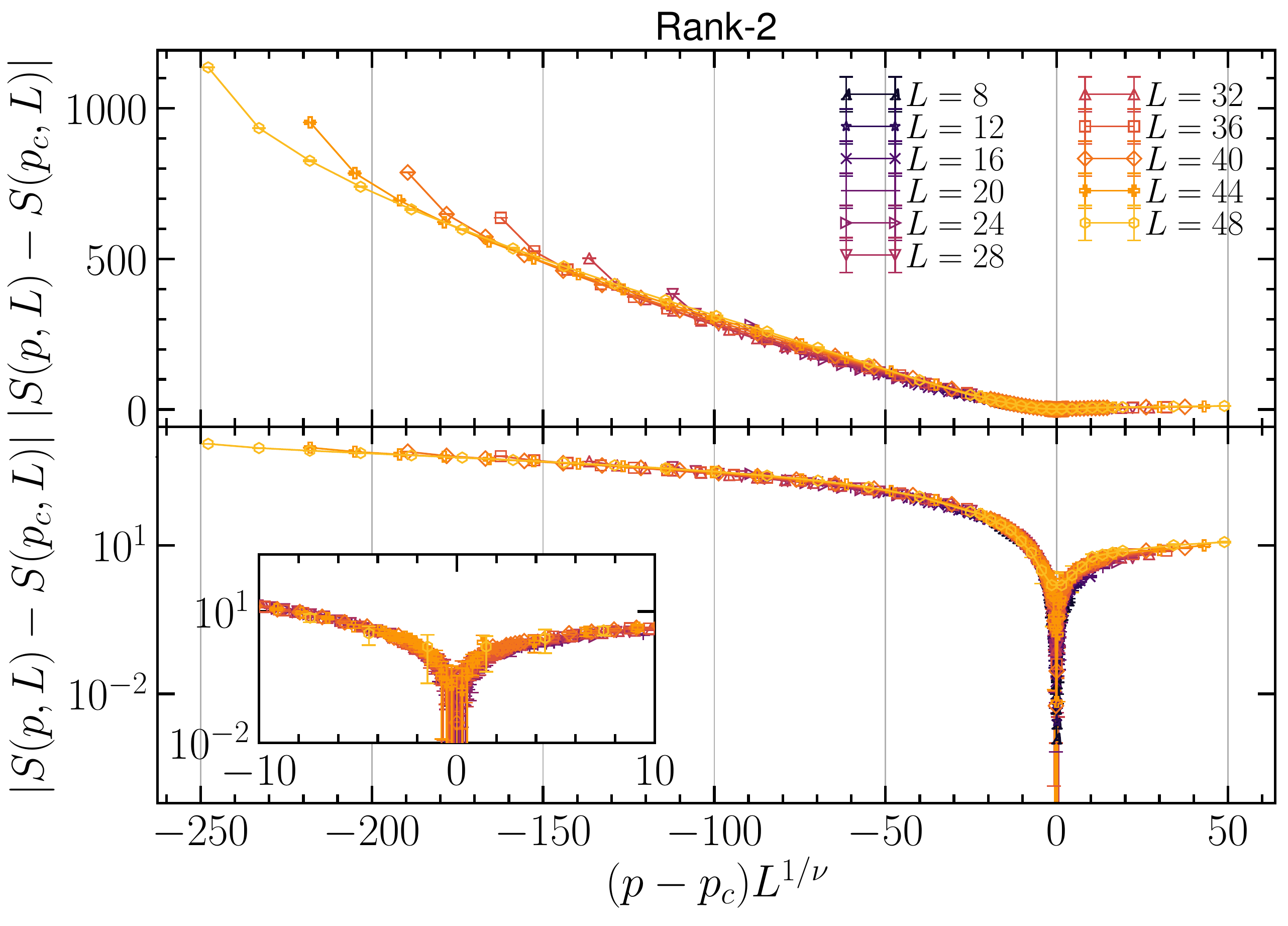}
	\caption{\label{fig3} Data collapse on the results of the finite size scaling in Fig.~\ref{fig2}. Rank-1 HRC results are presented on the upper panel, while the rank-2 HRC are on the lower one.}
\end{figure}

\begin{figure}[t!]
	\includegraphics[width=0.9\columnwidth]{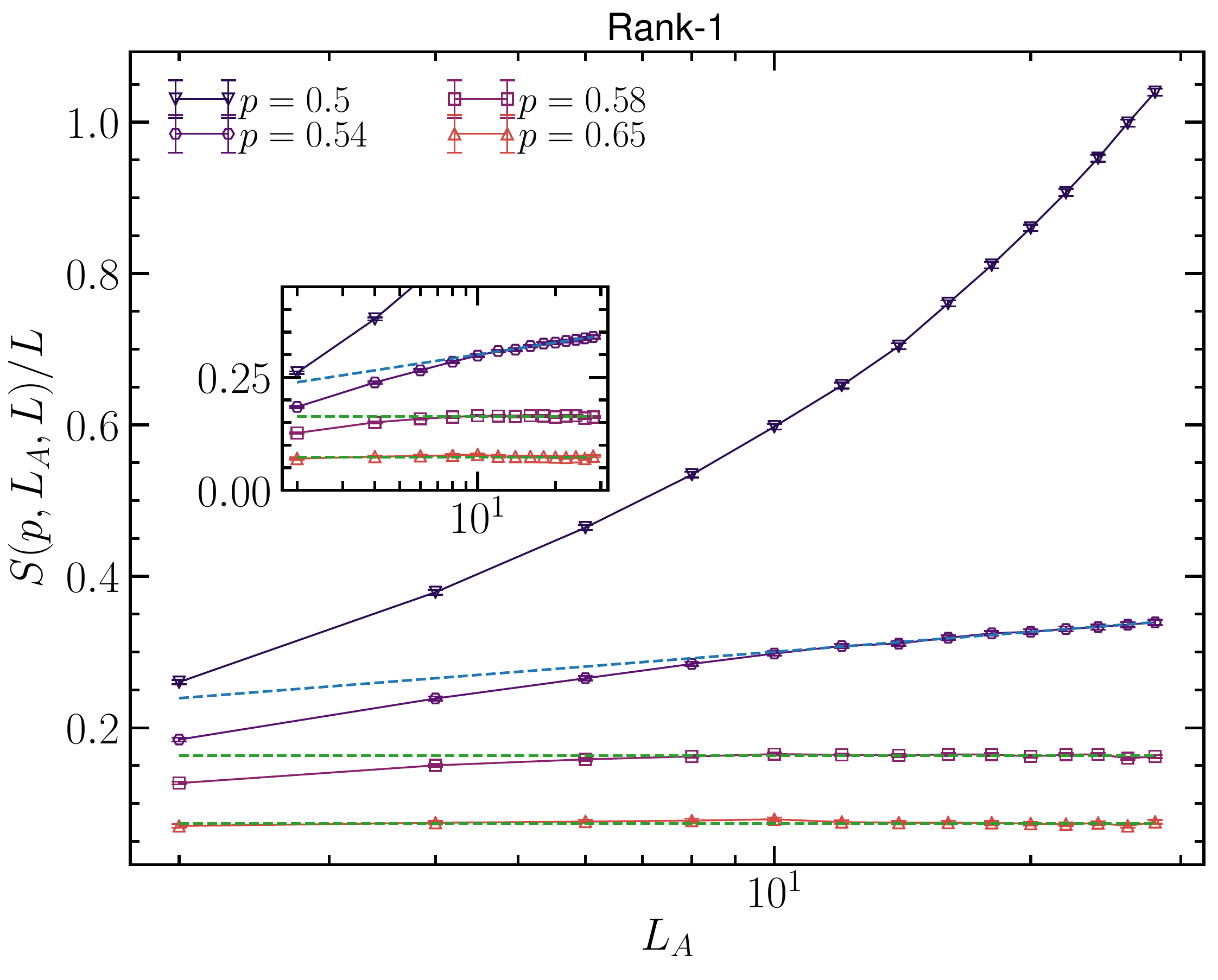}
	\includegraphics[width=0.9\columnwidth]{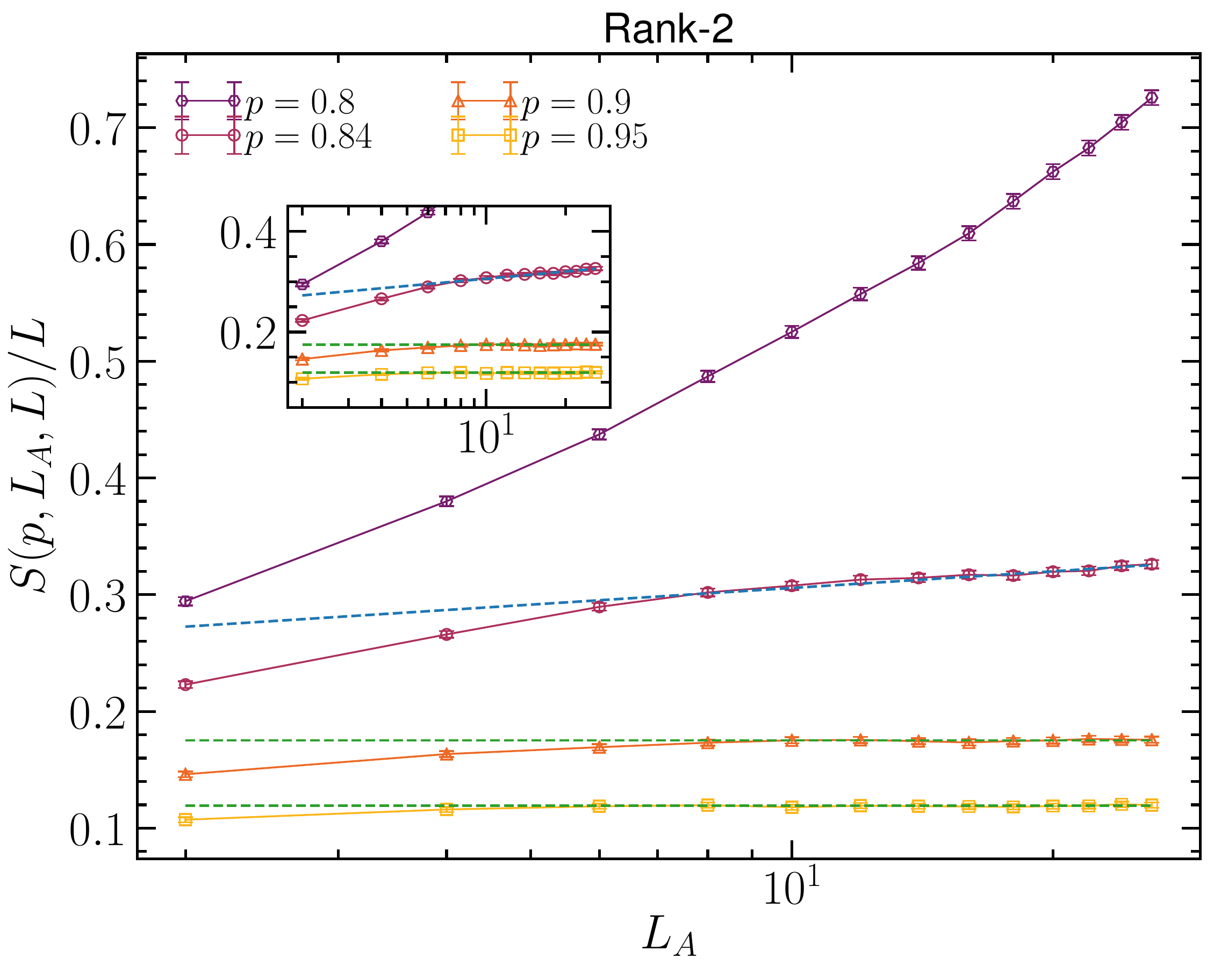}
	\caption{\label{fig4} Entanglement entropy for values around the criticality at $L=64$ for rank-1 (top panel) and rank-2 (bottom panel) HRC. The blue line is a fit for a logarithmic behavior, while the green ones are fit for an area-law. }
\end{figure}
\clearpage


\begin{thebibliography}{99}
%%%%%%%%%%%%%%%%%%%%%%%%%%%%%%%%%%%%%%%%%%%%%
\bibitem{Amico2007} 
L.~Amico, R.~Fazio, A.~Osterloh and V.~Vedral,
\newblock {\it Entanglement in many-body systems},
\newblock \href{http://dx.doi.org/10.1103/RevModPhys.80.517}{Rev. Mod. Phys. \textbf{80}, 517 (2008)}.

\bibitem{Calabrese2009R}
P.~Calabrese, J.~Cardy and B.~Doyon, 
\newblock {\it Entanglement entropy in extended quantum systems},
\newblock \href{http://dx.doi.org/10.1088/1751-8121/42/50/500301}{J. Phys. A  \textbf{42}, 500301 (2009)}.

\bibitem{Eisert2010}
J.~Eisert, M.~Cramer and M.~B.~Plenio,
\newblock {\it Area laws for the entanglement entropy},
\newblock \href{https://doi.org/10.1103/RevModPhys.82.277}{Rev. Mod. Phys. \textbf{82}, 277 (2010)}.

\bibitem{Laflorencie2015}
N.~Laflorencie,
\newblock {\it Quantum entanglement in condensed matter systems},
\newblock \href{http://dx.doi.org/10.1016/j.physrep.2016.06.008}{Phys. Rep.  \textbf{646}, 1 (2016)}.

\bibitem{Nandkishore2015}
R.~Nandkishore, and D.~A.~Huse,
\newblock \textit{Many-body localization and thermalization in quantum statistical mechanics},
\newblock \href{http://dx.doi.org/10.1146/annurev-conmatphys-031214-014726}{Annu. Rev. Condens. Matter Phys. \textbf{6}, 15 (2015)}.

\bibitem{Abanin2019}
D.~A.~Abanin, E.~Altman, I.~Bloch, and M.~Serbyn,
\newblock \textit{Many-body localization, thermalization, and entanglement},
\newblock \href{https://doi.org/10.1103/RevModPhys.91.021001}{Rev. Mod. Phys. \textbf{91}, 021001 (2019)}.

\bibitem{Smith2017}
A.~Smith, J.~Knolle, R.~Moessner, and D.~L.~Kovrizhin
\newblock \textit{Absence of Ergodicity without Quenched Disorder: From Quantum Disentangled Liquids to Many-Body Localization},
\newblock \href{https://doi.org/10.1103/PhysRevLett.119.176601}{Phys. Rev. Lett. \textbf{119}, 176601 (2017)}.

\bibitem{Brenes2018}
M.~Brenes, M.~Dalmonte, M.~Heyl, and A.~Scardicchio,
\newblock \textit{Many-body localization dynamics from gauge invariance},
\newblock \href{https://doi.org/10.1103/PhysRevLett.120.030601}{Phys. Rev. Lett. \textbf{120}, 030601 (2018)}.

\bibitem{Surace2020}
F.~M.~Surace, P.~P.~Mazza, G.~Giudici, A.~Lerose, A.~Gambassi, and M.~Dalmonte,
\newblock \textit{Lattice Gauge Theories and String Dynamics in Rydberg Atom Quantum Simulators},
\newblock \href{https://doi.org/10.1103/PhysRevX.10.021041}{Phys. Rev. X \textbf{10}, 021041 (2020)}.

\bibitem{Sala2020}
P.~Sala, T.~Rakovszky, R.~Verresen, M.~Knap, and F.~Pollmann,
\newblock \textit{Ergodicity Breaking Arising from Hilbert Space Fragmentation in Dipole-Conserving Hamiltonians},
\newblock \href{https://doi.org/10.1103/PhysRevX.10.011047}{Phys. Rev. X \textbf{10}, 011047 (2020)}.

\bibitem{Russomanno2020}
A.~Russomanno, S.~Notarnicola, F.~M.~Surace, R.~Fazio, M.~Dalmonte, and M.~Heyl,
\newblock \textit{Ergodicity Breaking Arising from Hilbert Space Fragmentation in Dipole-Conserving Hamiltonians},
\newblock \href{https://doi.org/10.1103/PhysRevResearch.2.012003}{Phys. Rev. Research \textbf{2}, 012003(R) (2020)}.

\bibitem{Schachenmayer2013}
J.~Schachenmayer, B.~P.~Lanyon, C.~F.~Roos, and A.~J.~Daley,
\newblock \textit{Entanglement Growth in Quench Dynamics with Variable Range Interactions},
\newblock \href{https://doi.org/10.1103/PhysRevX.3.031015}{Phys. Rev. X \textbf{3}, 031015 (2013)}.

\bibitem{Buyskikh2016} 
A.~S.~Buyskikh, M.~Fagotti, J.~Schachenmayer, F.~Essler, and A.~J.~Daley,
\newblock \textit{Entanglement growth and correlation spreading with variable-range interactions in spin and fermionic tunneling models},
\newblock \href{https://doi.org/10.1103/PhysRevA.93.053620}{Phys. Rev. A \textbf{93}, 053620 (2016)}.

\bibitem{Frerot2018}
I.~Fr\'erot, P.~Naldesi, and T.~Roscilde
\newblock \textit{Multispeed Prethermalization in Quantum Spin Models with Power-Law Decaying Interactions},
\newblock \href{https://doi.org/10.1103/PhysRevLett.120.050401}{Phys. Rev. Lett. \textbf{120}, 050401 (2018)}.

\bibitem{Liu2019}
F.~Liu, R.~Lundgren, P.~Titum, G.~Pagano, J.~Zhang, C.~Monroe, and A.~V.~Gorshkov,
\newblock \textit{Confined Quasiparticle Dynamics in Long-Range Interacting Quantum Spin Chains},
\newblock \href{https://doi.org/10.1103/PhysRevLett.122.150601}{Phys. Rev. Lett. \textbf{122}, 150601 (2019)}.

\bibitem{Lerose2020}
A.~Lerose, and S.~Pappalardi,
\newblock \textit{Origin of the slow growth of entanglement entropy in long-range interacting spin systems},
\newblock \href{https://doi.org/10.1103/PhysRevResearch.2.012041}{Phys. Rev. Research \textbf{2}, 012041(R) (2020)}.

\bibitem{Lerose2020B}
A.~Lerose, and S.~Pappalardi,
\newblock \textit{Bridging entanglement dynamics and chaos in semiclassical systems},
\newblock \href{https://arxiv.org/abs/2005.03670}{arXiv:2005.03670 (2020)}.

\bibitem{Calabrese2005}
P. Calabrese, and J. Cardy, 
\newblock {\it Evolution of Entanglement Entropy in One-Dimensional Systems},
\newblock \href{http://dx.doi.org/10.1088/1742-5468/2005/04/P04010}{J. Stat. Mech. (2005) P04010}.

\bibitem{Calabrese2007}
P.~Calabrese, and J.~Cardy,    
\newblock {\it Quantum quenches in extended systems},
\newblock \href{http://dx.doi.org/10.1088/1742-5468/2007/06/P06008}{J. Stat. Mech. \textbf{2007},  06008  (2007)}.

\bibitem{Chiara2006}
G.~De~Chiara, S.~Montangero, P.~Calabrese, and R.~Fazio,
\newblock {\it Entanglement entropy dynamics of Heisenberg chains},
\newblock \href{https://doi.org/10.1088/1742-5468/2006/03/P03001}{J. Stat. Mech. \textbf{2006}, 03001 (2006)}.

\bibitem{Rigol2008}
M.~Rigol, V.~Dunjko, and M.~Olshanii,
\newblock {\it Thermalization and its mechanism for generic isolated quantum systems},
\newblock \href{http://dx.doi.org/10.1038/nature06838}{Nature \textbf{452}, 854 (2008)}.

\bibitem{Vengalattore2011}
A.~Polkovnikov, K.~Sengupta, A.~Silva, and M.~Vengalattore,
\newblock {\it Nonequilibrium dynamics of closed interacting quantum systems},
\newblock \href{https://doi.org/10.1103/RevModPhys.83.863}{Rev. Mod. Phys. \textbf{83}, 863 (2011)}.

\bibitem{Kim2013}
H.~Kim, and D.~A.~Huse,
\newblock {\it Ballistic spreading of entanglement in a diffusive nonintegrable system},
\newblock \href{http://dx.doi.org/10.1103/PhysRevLett.111.127205}{Phys.  Rev.  Lett. \textbf{111},  127205  (2013)}.

\bibitem{Mezei2017}
M.~Mezei, and D.~Stanford,
\newblock {\it On entanglement spreading in chaotic quantum systems},
\newblock \href{http://dx.doi.org/10.1007/JHEP05(2017)065}{J. High Energy Phys. \textbf{2017}, 65 (2017)}.

\bibitem{Nahum2017}
A.~Nahum, J.~Ruhman, S.~Vijay, and J.~Haah,
\newblock \textit{Quantum entanglement growth under random unitary dynamics},
\newblock \href{https://doi.org/10.1103/PhysRevX.7.031016}{Phys. Rev. X \textbf{7}, 031016 (2017)}.

\bibitem{Bertini2019}
B.~Bertini, P.~Kos, and T.~Prosen,
\newblock \textit{Entanglement Spreading in a Minimal Model of Maximal Many-Body Quantum Chaos},
\newblock \href{https://doi.org/10.1103/PhysRevX.9.021033}{Phys. Rev. X \textbf{9}, 021033 (2019)}.

\bibitem{Piroli2020}
L.~Piroli, B.~Bertini, J.~I.~Cirac, and T.~Prosen,
\newblock \textit{Exact dynamics in dual-unitary quantum circuits},
\newblock \href{https://doi.org/10.1103/PhysRevB.101.094304}{Phys. Rev. B \textbf{101}, 094304 (2020)}.

\bibitem{Bera2020}
A.~Bera, and S.~S.~Roy,
\newblock \textit{Growth of genuine multipartite entanglement in random unitary circuits},
\newblock \href{https://arxiv.org/abs/2003.12546}{arXiv: 2003.12546 (2020)}.

\bibitem{Daley2012}
A.~J.~Daley, H.~Pichler, J.~Schachenmayer, and P.~Zoller,
\newblock {\it Measuring Entanglement Growth in Quench Dynamics of Bosons in an Optical Lattice},
\newblock \href{https://doi.org/10.1103/PhysRevLett.109.020505}{Phys. Rev. Lett. \textbf{109}, 020505 (2012)}.

\bibitem{Islam2015}
R.~Islam, R.~Ma, P.~M. Preiss, M.~E. Tai, A.~Lukin, M.~Rispoli and M.~Greiner,
\newblock {\it Measuring entanglement entropy in a quantum many-body system},
\newblock \href{http://dx.doi.org/10.1038/nature15750}{Nature \textbf{528}, 77 (2015)}.

\bibitem{Kaufman2016}
A. M.~Kaufman, M.E.~Tai, A.~Lukin, M.~Rispoli, R.~Schittko, P. M.~Preiss and M.~Greiner,
\newblock {\it Quantum thermalization through entanglement in an isolated  many-body system},
\newblock \href{http://dx.doi.org/10.1126/science.aaf6725}{Science \textbf{353}, 764 (2016)}. 
 
\bibitem{Elben2018}
A.~Elben, B.~Vermersch, M.~Dalmonte, J.I.~Cirac and P.~Zoller, 
\newblock {\it R\'enyi Entropies from Random Quenches in Atomic Hubbard and Spin Models}, 
\newblock \href{http://dx.doi.org/10.1103/PhysRevLett.120.050406}{Phys. Rev. Lett. \textbf{120}, 050406 (2018)}.

\bibitem{Brydges2019}
T.~Brydges, A.~Elben, P.~Jurcevic, B.~Vermersch, C.~Maier, B.P.~Lanyon, P.~Zoller, R.~Blatt and C.F.~Roos,
\newblock {\it Probing R\'enyi entanglement entropy via randomized measurements},
\newblock \href{https://doi.org/10.1126/science.aau4963}{Science {\bf 364}, 6437 (2019)}.

\bibitem{Lukin2019}
A.~Lukin, M.~Rispoli, R.~Schittko, M.E.~Tai, A.M.~Kaufman, S.~Choi, V.~Khemani, J.~Leonard and M.Z.~Greiner,
\newblock {\it Probing entanglement in a many-body localized system},
\newblock \href{http://dx.doi.org/10.1126/science.aau0818}{Science \textbf{364}, 6437 (2019)}.

\bibitem{Cao2019}
X.~Cao, A.~Tilloy, and A.~De~Luca,
\newblock \textit{Entanglement in a fermion chain under continuous monitoring},
\newblock \href{http://dx.doi.org/10.21468/SciPostPhys.7.2.024}{SciPost Phys. \textbf{7}, 024 (2019)}.

\bibitem{Chan2019}
A.~Chan, R.~M.~Nandkishore, M.~Pretko, and G.~Smith,
\newblock \textit{Unitary-projective entanglement dynamics},
\newblock \href{https://doi.org/10.1103/PhysRevB.99.224307}{Phys. Rev. B \textbf{99}, 224307 (2019)}.

\bibitem{Li2019}
Y.~Li, X.~Chen, and M.~P.~A.~Fisher,
\newblock \textit{Measurement-driven entanglement transition in hybrid quantum circuits},
\newblock \href{https://doi.org/10.1103/PhysRevB.100.134306}{Phys. Rev. B \textbf{100}, 134306 (2019)}.

\bibitem{Li2018}
Y.~Li, X.~Chen, and M.~P.~A.~Fisher,
\newblock \textit{Quantum Zeno effect and the many-body entanglement transition},
\newblock \href{https://doi.org/10.1103/PhysRevB.98.205136}{ Phys. Rev. B \textbf{98}, 205136 (2018)}.

\bibitem{Skinner2019}
B.~Skinner, J.~Ruhman, and A.~Nahum,
\newblock \textit{Measurement-induced phase transitions in the dynamics of entanglement},
\newblock \href{https://doi.org/10.1103/PhysRevX.9.031009}{ Phys. Rev. X \textbf{9}, 031009 (2019)}.

\bibitem{Choi2019} 
S.~Choi, Y.~Bao, X.~L.~Qi, and E.~Altman,
\newblock \textit{Quantum error correction in scrambling dynamics and measurement induced phase transition},
\newblock \href{https://arxiv.org/abs/1903.05124}{arxiv: 1903.05124 (2019)}.

\bibitem{Bao2019}
Y.~Bao, S.~Choi, and E.~Altman,
\newblock \textit{Theory of the phase transition in random unitary circuits with measurements},
\newblock \href{https://doi.org/10.1103/PhysRevB.101.104301}{Phys. Rev. B \textbf{101}, 104301  (2020)}.

\bibitem{Gullans2019}
M.~J.~Gullans, and D.~A.~Huse,
\newblock \textit{Dynamical purification phase transition induced by quantum measurements},
\newblock \href{https://arxiv.org/abs/1905.05195}{arXiv: 1905.05195 (2019)}.

\bibitem{Gullans2019B}
M.~J.~Gullans, and D.~A.~Huse,
\newblock \textit{Scalable probes of measurement-induced criticality},
\newblock \href{https://arxiv.org/abs/1910.00020}{arXiv: 1910.00020 (2019)}.

\bibitem{Szyniszewski2019}
M.~Szyniszewski, A.~Romito, and H.~Schomerus,
\newblock \textit{Entanglement transition from variable-strength weak measurements},
\newblock \href{https://doi.org/10.1103/PhysRevB.100.064204}{ Phys. Rev. B \textbf{100}, 064204 (2019)}.

\bibitem{Tang2019}
Q.~Tang, and W.~Zhu,
\newblock \textit{Measurement-induced phase transition: A case study in the non-integrable model by density-matrix renormalization group calculations},
\newblock \href{https://doi.org/10.1103/PhysRevResearch.2.013022}{Phys. Rev. Research \textbf{2}, 013022 (2020)}.

\bibitem{Jian2019}
C.~M.~Jian, Y.~Z.~You, R.~Vasseur, and A.~W.~W.~Ludwig,
\newblock \textit{Measurement-induced criticality in random quantum circuits},
\newblock \href{https://doi.org/10.1103/PhysRevB.101.104302}{Phys. Rev. B \textbf{101}, 104302 (2020)}.

\bibitem{Zabalo2019}
A.~Zabalo, M.~J.~Gullans, J.~H.~Wilson, S.~Gopalakrishnan, D.~A.~Huse, and J.~H.~Pixley,
\newblock \textit{Critical properties of the measurement-induced transition in random quantum circuits},
\newblock \href{https://doi.org/10.1103/PhysRevB.101.060301}{Phys. Rev. B \textbf{101}, 060301(R) (2020)}.

\bibitem{Zhang2020}
L.~Zhang, J.~A.~Reyes, S.~Kourtis, C.~Chamon, E.~R.~Mucciolo, and A.~E.~Ruckenstein,
\newblock \textit{Non-universal Entanglement Level Statistics in Projection-driven Quantum Circuits},
\newblock \href{https://doi.org/10.1103/PhysRevB.101.235104}{Phys. Rev. B \textbf{101}, 235104 (2020)}.

\bibitem{Kuo2019}
W.-T.~Kuo, A.~A.~Akhtar, D.~P.~Arovas, and Y.~-Z.~You,
\newblock \textit{Markovian Entanglement Dynamics under Locally Scrambled Quantum Evolution},
\newblock \href{https://doi.org/10.1103/PhysRevB.101.224202}{Phys. Rev. B \textbf{101}, 224202 (2020)}.

\bibitem{Gebhart2019}
V.~Gebhart, K.~Snizhko, T.~Wellens, A.~Buchleitner, A.~Romito, and Y.~Gefen,
\newblock \textit{Topological transition in measurement-induced geometric phases},
\newblock \href{https://doi.org/10.1073/pnas.1911620117}{PNAS \textbf{117}, 5706 (2019)}.

\bibitem{Snizhko2020}
K.~Snizhko, P.~Kumar, and A.~Romito,
\newblock \textit{The Quantum Zeno effect appears in stages},
\newblock \href{https://arxiv.org/abs/2003.10476}{arXiv: 2003.10476 (2020)}.

\bibitem{Goto2020}
S.~Goto, and I.~Danshita,
\newblock \textit{Measurement-Induced Transitions of the Entanglement Scaling Law in Ultracold Gases with Controllable Dissipation},
\newblock \href{https://arxiv.org/abs/2001.03400}{arXiv: 2001.03400 (2020)}.

\bibitem{Rossini2020}
D.~Rossini, and E.~Vicari,
\newblock \textit{Measurement-induced dynamics of many-body systems at quantum criticality},
\newblock \href{https://arxiv.org/abs/2001.11501}{arXiv: 2001.11501 (2020)}.

\bibitem{Fan2020}
R.~Fan, S.~Vijay, A.~Vishwanath, and Y.~-Z.~You,
\newblock \textit{Self-Organized Error Correction in Random Unitary Circuits with Measurement},
\newblock \href{https://arxiv.org/abs/2002.12385}{arXiv: 2002.12385 (2020)}.


\bibitem{Shtanko2020}
O.~Shtanko, Y.~A.~Kharkov, L.~P.~Garc\'ia-Pintos, and A.~V.~Gorshkov,
\newblock \textit{Classical Models of Entanglement in Monitored Random Circuits},
\newblock \href{https://arxiv.org/abs/2004.06736}{arXiv: 2004.06736 (2020)}.

\bibitem{Piqueres2020}
J.~Lopez-Piqueres, B.~Ware, and R.~Vasseur,
\newblock \textit{Mean-field theory of entanglement transitions from random tree tensor networks},
\newblock \href{https://arxiv.org/abs/2003.01138}{arXiv: 2003.01138 (2020)}.

\bibitem{Szyniszewski2020}
M.~Szyniszewski, A.~Romito, and H.~Schomerus,
\newblock \textit{Universality of entanglement transitions from stroboscopic to continuous measurements},
\newblock \href{https://arxiv.org/abs/2005.01863}{arXiv: 2005.01863 (2020)}.

\bibitem{Fuji2020}
Y.~Fuji, and Y.~Ashida,
\newblock \textit{Measurement-induced quantum criticality under continuous monitoring},
\newblock \href{https://arxiv.org/abs/2004.11957}{arXiv: 2004.11957 (2020)}.

\bibitem{Chen2020}
X.~Chen, Y.~Li, M.~P.~A.~Fisher, and A.~Lucas,
\newblock \textit{Emergent conformal symmetry in non-unitary random dynamics of free fermions},
\newblock \href{https://arxiv.org/abs/2004.09577}{arXiv: 2004.09577 (2020)}.

\bibitem{Li2020}
Y.~Li, X.~Chen, A.~W.~W.~Ludwig, and M.~P.~A.~Fisher,
\newblock \textit{Conformal invariance and quantum non-locality in hybrid quantum circuits},
\newblock \href{https://arxiv.org/abs/2003.12721}{arXiv: 2003.12721 (2020)}.

\bibitem{Lavasani2020}
A.~Lavasani, Y.~Alavirad, and M.~Barkeshli,
\newblock \textit{Measurement-induced topological entanglement transitions in symmetric random quantum circuits},
\newblock \href{https://arxiv.org/abs/2004.07243}{arXiv: 2004.07243 (2020)}.

\bibitem{Ippoliti2020}
M.~Ippoliti, M.~J.~Gullans, S.~Gopalakrishnan, D.~A.~Huse, and V.~Khemani,
\newblock \textit{Entanglement phase transitions in measurement-only dynamics},
\newblock \href{https://arxiv.org/abs/2004.09560}{arXiv: 2004.09560 (2020)}.

\bibitem{Sang2020}
S.~Sang, and T.~H.~Hsieh,
\newblock \textit{Measurement Protected Quantum Phases},
\newblock \href{https://arxiv.org/abs/2004.09509}{arXiv: 2004.09509 (2020)}.

\bibitem{Lang2020}
N.~Lang, and H.~P.~B\"uchler,
\newblock \textit{Entanglement Transition in the Projective Transverse Field Ising Model},
\newblock \href{https://arxiv.org/abs/2006.09748}{arXiv: 2006.09748 (2020)}.

\bibitem{Lunt2020}
O.~Lunt, and A.~Pal,
\newblock \textit{Measurement-induced entanglement transitions in many-body localized systems},
\newblock \href{https://arxiv.org/abs/2005.13603}{arXiv: 2005.13603 (2020)}.

\bibitem{Wolf2006}
M.~M.~Wolf,
\newblock \textit{Violation of the Entropic Area Law for Fermions},
\newblock \href{https://doi.org/10.1103/PhysRevLett.96.010404}{Phys. Rev. Lett. \textbf{96}, 010404 (2006)}.

\bibitem{Gioev2006}
D.~Gioev, and I.~Klich
\newblock \textit{Entanglement Entropy of Fermions in Any Dimension and the Widom Conjecture},
\newblock \href{https://doi.org/10.1103/PhysRevLett.96.100503}{Phys. Rev. Lett. \textbf{96}, 100503 (2006)}.

\bibitem{motrunich2008comparative}
O. I. Motrunich and A. Vishwanath, 
\newblock \textit{Comparative study of Higgs transition in one-component and two-component lattice superconductor models},
\newblock \href{https://arxiv.org/abs/0805.1494}{arXiv:0805.1494 (2008)}.

\bibitem{Swingle2010}
B.~Swingle,
\newblock \textit{Entanglement Entropy and the Fermi Surface},
\newblock \href{https://doi.org/10.1103/PhysRevLett.105.050502}{Phys. Rev. Lett. \textbf{105}, 050502 (2010)}.

\bibitem{Zhang2011}
Y.~Zhang, T.~Grover, and A.~Vishwanath,
\newblock \textit{Entanglement Entropy of Critical Spin Liquids},
\newblock \href{hhttps://doi.org/10.1103/PhysRevLett.107.067202}{Phys. Rev. Lett. \textbf{107}, 067202 (2011)}.

\bibitem{Potter2014}
A.~C.~Potter,
\newblock \textit{Boundary-law scaling of entanglement entropy in diffusive metals},
\newblock \href{https://arxiv.org/abs/1408.1094}{arXiv:1408.1094 (2014)}.

\bibitem{Pouranvari2015} 
M.~Pouranvari, Y.~Zhang, and K.~Yang,
\newblock \textit{Entanglement Area Law in Disordered Free Fermion Anderson Model in One, Two, and Three Dimensions},
\newblock \href{https://doi.org/10.1155/2015/397630}{Adv. Condens. Matter Phys. \textbf{2015}, 397630 (2015)}.

\bibitem{Swingle2016}
B.~Swingle, and J.~McGreevy,
\newblock \textit{Area law for gapless states from local entanglement thermodynamics},
\newblock \href{httpsL//doi.org/10.1103/PhysRevB.93.205120}{Phys. Rev. B \textbf{93}, 205120 (2016)}.

\bibitem{foot1}
\newblock{The choice of low-entanglement state is unimportant for our purposes, as the stationary regime is independent of the initial state. See further below.}

\bibitem{foot2}
\newblock{Sometimes, those realizations are called trajectories by analogy with their continuous-time counterpart. We prefer to keep the wording distinct as quantum trajectories are representative of a very different class of dynamics, where projections are inserted within a {\it non-unitary} dynamics.}

\bibitem{Nielsen2012}
M.~A.~Nielsen, and I.~L.~Chuang,
\newblock \textit{Quantum Computation and Quantum Information},
\newblock \href{https://doi.org/10.1017/CBO9780511976667}{10th anniversary edition, CUP, Cambridge (2010).}.

\bibitem{Gottesman1998}
D.~Gottesman,
\newblock \textit{The Heisenberg Representation of Quantum Computers},
\newblock \href{https://arxiv.org/abs/quant-ph/9807006v1}{arXiv: 9807006 (1998)}.

\bibitem{Aaronson2004}
S.~Aaronson, and D.~Gottesman,
\newblock \textit{Improved simulation of stabilizer circuits},
\newblock \href{https://doi.org/10.1103/PhysRevA.70.052328}{Phys.~Rev.~A  \textbf{70}, 052328 (2014)}.

\bibitem{Hamma2004}
A.~Hamma, R.~Ionicioiu, and P.~Zanardi,
\newblock \textit{Ground state entanglement and geometric entropy in the Kitaev's model},
\newblock \href{https://doi.org/10.1016/j.physleta.2005.01.060}{Phys.~Lett.~A \textbf{337}, 22 (2005)}.

\bibitem{Hamma2005}
A.~Hamma, R.~Ionicioiu, and P.~Zanardi,
\newblock \textit{Bipartite entanglement and entropic boundary law in lattice spin systems},
\newblock \href{https://doi.org/10.1103/PhysRevA.71.022315}{Phys.~Rev.~A \textbf{71}, 022315 (2005)}.

\bibitem{Koenig2014}
R.~Koenig, and J.~A.~Smolin,
\newblock \textit{How to efficiently select an arbitrary Clifford group element},
\newblock \href{https://doi.org/10.1063/1.4903507}{J.~Math.~Phys. \textbf{55}, 122202 (2014)}.

\bibitem{Khemani2018}
V.~Khemani, A.~Vishwanath, D.~A.~Huse,
\newblock \textit{Operator spreading and the emergence of dissipative hydrodynamics under unitary evolution with conservation laws},
\newblock \href{https://doi.org/10.1103/PhysRevX.8.031057}{Phys. Rev. X \textbf{8}, 031057 (2018)}.

\bibitem{Nahum2018B}
A.~Nahum, S.~Vijay, and J.~Haah,
\newblock \textit{Operator Spreading in Random Unitary Circuits},
\newblock \href{https://doi.org/10.1103/PhysRevX.8.021014}{Phys. Rev. X \textbf{8}, 021014 (2018)}.

\bibitem{Pollmann2018}
C.~W. von~Keyserlingk, T.~Rakovszky, F.~Pollmann, and S.~L.~Sondhi,
\newblock \textit{Operator Hydrodynamics, OTOCs, and Entanglement Growth in Systems without Conservation Laws},
\newblock \href{https://doi.org/10.1103/PhysRevX.8.021013}{Phys. Rev. X \textbf{8}, 021013 (2018)}.

\bibitem{Grover2015}
T.~Grover, Y.~Zhang, A.~Vishwanath,
\newblock \textit{Entanglement entropy as a portal to the physics of quantum spin liquids},
\newblock \href{https://doi.org/10.1088/1367-2630/15/2/025002}{New J. Phys. \textbf{15}, 025002 (2013).}

\bibitem{Turkeshi2020}
X.~Turkeshi et al., 
\newblock \textit{In progress}.

\end{thebibliography}
\end{document}